\documentclass[lettersize,journal]{IEEEtran}
\usepackage{amsmath,amsfonts}
\usepackage{algorithmic}
\usepackage{array}
\usepackage[font=footnotesize, labelsep=period]{caption}
\usepackage{textcomp}
\usepackage{stfloats}
\usepackage{url}
\usepackage{verbatim}
\usepackage{graphicx}
\def\BibTeX{{\rm B\kern-.05em{\sc i\kern-.025em b}\kern-.08em
    T\kern-.1667em\lower.7ex\hbox{E}\kern-.125emX}}
\usepackage{balance}

\usepackage{latexsym}
\usepackage{graphicx}
\usepackage{amsfonts,amssymb,amsmath}
\usepackage{hyperref}
\hypersetup{hidelinks}
\def\BibTeX{{\rm B\kern-.05em{\sc i\kern-.025em b}\kern-.08em T\kern-.1667em\lower.7ex\hbox{E}\kern-.125emX}}

\usepackage{pifont}

\usepackage{algorithm}

\usepackage{xcolor} 
\usepackage{colortbl}

\usepackage{cite}

\newfont{\bbb}{msbm10 scaled 500}

\newfont{\bb}{msbm10 scaled 1100}

\newcommand{\RR}{\mbox{\bb R}}

\newcommand{\av}{{\bf a}}

\newcommand{\cv}{{\bf c}}

\newcommand{\ev}{{\bf e}}

\newcommand{\sv}{{\bf s}}

\newcommand{\xv}{{\bf x}}

\newcommand{\zv}{{\bf z}}

\newcommand{\Am}{{\bf A}}
\newcommand{\Bm}{{\bf B}}

\newcommand{\Dm}{{\bf D}}

\newcommand{\Gm}{{\bf G}}
\newcommand{\Hm}{{\bf H}}
\newcommand{\Id}{{\bf I}}

\newcommand{\Pm}{{\bf P}}
\newcommand{\Qm}{{\bf Q}}

\newcommand{\Wm}{{\bf W}}
\newcommand{\Vm}{{\bf V}}

\newcommand{\Ym}{{\bf Y}}

\newcommand{\deltav}{\hbox{\boldmath$\delta$}}

\newcommand{\psiv}{\hbox{\boldmath$\psi$}}
\newcommand{\thetav}{\hbox{\boldmath$\theta$}}

\newcommand{\omegav}{\hbox{\boldmath$\omega$}}

\newcommand{\diag}{{\hbox{diag}}}

\renewcommand{\arg}{{\hbox{arg}}}

\newcommand{\defines}{{\,\,\stackrel{\scriptscriptstyle \bigtriangleup}{=}\,\,}}

\usepackage{bbm}
\usepackage{subcaption}
\usepackage{tabularx}
\usepackage{adjustbox}

\usepackage{multirow}
\usepackage{verbatim}
\usepackage{color}
\definecolor{orange}{RGB}{222, 125, 44}
\usepackage{mathrsfs}
\usepackage{float}
\usepackage{bm}
\usepackage{booktabs}
\usepackage{diagbox}
\usepackage{makecell}
\usepackage{bbding}
\usepackage{amssymb}
\usepackage{nomencl}
\makenomenclature
\usepackage{etoolbox}

\renewcommand\arraystretch{2}

\newcommand{\beqa}{\begin{eqnarray}}
\newcommand{\eeqa}{\end{eqnarray}}

\newcommand{\chen}[1]{{\color{blue} #1}}

\begin{document}

\title{Moving Target Defense Against Adversarial False Data Injection Attacks In Power Grids}

\author{Yexiang Chen~\IEEEmembership{Member, IEEE}, Subhash Lakshminarayana~\IEEEmembership{Senior Member, IEEE}, and H. Vincent Poor~\IEEEmembership{Fellow, IEEE}
\thanks{ Yexiang Chen and Subhash Lakshminarayana (corresponding author) are with School of Engineering, University of Warwick, UK (e-mail: \{yexiang.chen, subhash.lakshminarayana\}@warwick.ac.uk). H. Vincent Poor is with Department of Electrical and Computer Engineering, Princeton University, Princeton, NJ 08544, USA (e-mail: poor@princeton.edu).}
}

\markboth{Manuscript submitted for IEEE Internet of Things Journal}%
{Moving Target Defense Against Adversarial False Data Injection Attacks In Power Grids}

\IEEEaftertitletext{\vspace{-2.2\baselineskip}}
\maketitle

\begin{abstract}
Machine learning (ML)-based detectors have been shown to be effective in detecting stealthy false data injection attacks (FDIAs) that can bypass conventional bad data detectors (BDDs) in power systems. However, ML models are also vulnerable to adversarial attacks. A sophisticated perturbation signal added to the original BDD-bypassing FDIA can conceal the attack from ML-based detectors. 
In this paper, we develop a moving target defense (MTD) strategy to defend against adversarial FDIAs in power grids. We first develop an MTD-strengthened deep neural network (DNN) model, which deploys a pool of DNN models rather than a single static model that cooperate to detect the adversarial attack jointly. The MTD model pool introduces randomness to the ML model's decision boundary, thereby making the adversarial attacks detectable. Furthermore, to increase the effectiveness of the MTD strategy and reduce the computational costs associated with developing the MTD model pool, we combine this approach with the physics-based MTD, which involves dynamically perturbing the transmission line reactance and retraining the DNN-based detector to adapt to the new system topology. Simulations conducted on IEEE test bus systems demonstrate that the MTD-strengthened DNN achieves up to 94.2\% accuracy in detecting adversarial FDIAs. When combined with a physics-based MTD, the detection accuracy surpasses 99\%, while significantly reducing the computational costs of updating the DNN models. This approach requires only moderate perturbations to transmission line reactances, resulting in minimal increases in OPF cost.

\end{abstract}


%\vspace{-3mm}
\section{Introduction} \label{INTRODUCTION}

\IEEEPARstart{T}{he} vulnerability of power grid state estimation (SE) to false data injection attacks (FDIAs) has been a well-studied topic over the last decade \cite{liu2011false, bobba2010detecting, dan2010stealth, ozay2015machine, LakshminarayanaCostBenefit}. Early defence approaches to defend against bad data detector (BDD)-bypassing FDIAs included solutions such as carefully protecting a subset of sensors (e.g., via hardware updates such as using tamper-proof and encryption-enabled PLCs) or independently verifying a subset of strategically selected state variables using phase measurement units (PMUs) \cite{bobba2010detecting, dan2010stealth} in order to prevent the attacker from crafting BDD-bypassing FDIAs. However, these solutions incur high capital costs in terms of infrastructure upgrades (e.g., enabling encryption). 

 % Several approaches have been proposed to defend against the threat of bad data detector (BDD)-bypassing FDIAs. 
 
% Subsequent research has also shown that PMUs themselves are at risk of suffering from  FDIAs, such as through spoofing on GPS receivers\cite{Daniel2012}.

% Power grids are witnessing growing penetration of information and communication technologies (ICTs) to automate their operations. However, the penetration of ICTs also makes the grid vulnerable to cyber threats. In particular, , used for computing estimates of the power system state using the measurements taken from various sensors and devices, plays an important role, such as computing the optimal power flow, contingency analysis, etc. However, SE has been shown to be vulnerable to false data injection attacks (FDIAs) \cite{liu2011false}. Specifically, the FDIAs crafted using detailed knowledge of a power grid's topology and reactance settings of its transmission line can bypass the detection of bad data detectors (BDDs) and remain stealthy. Undetected FDIAs can have severe consequences, e.g., trips of transmission line breakers and unsafe frequency excursions\cite{lakshminarayana2017optimal}. 

% As a data-driven technique, the DNN-based detector is cost-effective and can be implemented in power system operation centres without requiring additional equipment or infrastructure upgrades. 

% , particularly the deep learning (DL) approach such as deep neural network (DNN), 

To defend against stealthy FDIAs, there is a growing interest in applying machine learning (ML) to detect BDD-bypassing FDIAs \cite{ozay2015machine, he2017real, ZhangSemi, ZhangML2020, wang2020locational, abdi2024role}. ML models are trained offline using large amounts of measurement data to learn the inconsistencies introduced by FDIAs, and are then able to provide accurate online identification on attack existence and localization. Despite the effectiveness of ML models, they are vulnerable to \emph{adversarial attacks} \cite{goodfellow2015explaining}. The basic principle behind designing adversarial attacks involves introducing carefully crafted perturbations to input data exploiting the model's sensitivity to imperceptible changes, causing it to make incorrect predictions while appearing nearly indistinguishable from the original input.
Adversarial FDIAs against DNN-based detectors in power grids have garnered increasing attention in recent studies \cite{huang2022adversarial, li2023towards, AFDIA2021, Tian2022target}. 
A key difference between adversarial attacks in power grids and in other domains such as image processing is that the attack must simultaneously bypass the detection from both the BDD as well as DNN-based detection \cite{AFDIA2021}.

Defending against adversarial attacks is a challenging problem. 
State-of-the-art methods to counter adversarial attacks on DNNs developed include adversarial training \cite{kurakin2016adversarial}, applying data transformation layers \cite{xie2017adversarial}, gradient masking techniques \cite{buckman2018thermometer}, etc. These static defense techniques have also been applied in the context of power grids \cite{li2023towards, GuihaiAdv2021}. However, these methods still have limitations. Adversarial training can reduce the model's performance on clean data, making it hard to balance robustness and generalization. Although models using gradient masking may resist specific perturbations encountered during training, they might still be vulnerable to new attack strategies. Moreover, these defenses are less effective against adaptive attackers who can learn the defense mechanisms, such as the algorithm used for generating adversarial examples \cite{athalye2018obfuscated, Zhong2021Transferable}.

% , Research has shown that even advanced defenses can be circumvented by adaptive attackers who adjust their techniques based on the specific defenses used \cite{athalye2018obfuscated, Zhong2021Transferable}.

%****
% Researchers in \cite{li2023towards} propose an adversarial-resilient DNN detection framework that incorporates random input padding to prevent attackers from successfully launching adversarial FDIAs.
%****

%Adaptive attackers can develop new strategies that exploit weaknesses not addressed during training, reverse transformations to counteract data alterations, bypass gradient masking by finding alternative methods for estimating gradients, and enhance transferability of adversarial attacks \cite{Zhong2021Transferable}. 

% , the data transformation process, and gradient masking techniques

Recently, a novel defense strategy known as moving target defense (MTD), characterized by its proactive and dynamic nature, has demonstrated its effectiveness in thwarting knowledgeable attackers. The fundamental concept behind MTD involves introducing periodic changes to the system in order to invalidate the knowledge that the attackers need to launch stealthy FDIAs. For example, in power grid applications, the knowledge of the Jacobian matrix is necessary to launch BDD-bypassing attacks. In this context, MTD design based on periodically perturbing transmission line reactance using physical devices, such as Distributed Flexible AC Transmission system (D-FACTS), has received significant attention \cite{LakshminarayanaCostBenefit, Lakshminarayana2021, xu2022blending, liu2020optimal, LiuMTD2018, xu2021robust, liu2022, ZhangMTD2020, CuiHidden2021, LiuMCMTD2021}. We refer to such MTD as \emph{physics-based MTD } as it involves perturbing the physical system. While significant research has been conducted on this topic (see Section II-B for more details), most of the works focus on designing MTD against BDD-bypassing attacks only. They are not designed to
counter the specific threat of adversarial perturbations that can bypass both the BDD and the DNN-based detector. Our results show that while
the physics-based MTD approach can detect adversarial attacks, they require large
reactance perturbations, which also incur significant operational costs. 
%This approach although provides some resilience against adversarial FDIAs, does not directly translate to defending against them. 

% To the best of our knowledge, there are no works in power grid context that consider the detection of stealthy adversarial attacks that bypass both the BDD and the ML-based detectors. 

%Furthermore, physics-based MTD incurs unavoidable operational costs due to the requirement of \emph{moving} the system away from its optimal settings for cyber defense purposes \cite{LakshminarayanaCostBenefit}. 

%\arrayrulecolor{blue}

\begin{table*}[!h]
\caption{Summarization of the paper contributions with respect to the existing literature}
\resizebox{0.99\textwidth}{!}{
    \begin{tabular}{|>{\centering\arraybackslash}p{4cm}|c|c|c|>{\centering\arraybackslash}p{4.5cm}|>{\centering\arraybackslash}p{7cm}|}
         \hline
         \multirow{2}{*}{Detection Approach} & \multicolumn{3}{|c|}{Attack Categories} & \multirow{2}{*}{Requirements for Implementation} & \multirow{2}{*}{Description} \\
         \cline{2-4}
         & Random Attack & \makecell{BDD-bypassing \\ FDIA} & Adversarial FDIA & & \\
         \hline
         BDD \cite{liu2011false} & \checkmark & \ding{55} & \ding{55} & Computational resources (BDD) &\makecell[l]{\parbox{7cm}{\vspace{0cm} \raggedright BDD can be bypassed by sophisticated FDIAs.}}\\
         \hline
         DNN-based Detector \cite{ozay2015machine, he2017real, ZhangSemi, ZhangML2020, wang2020locational} & \checkmark & \checkmark & \ding{55} & \makecell{ Computational resources \\ (single DNN model) } &\makecell[l]{\parbox{7cm}{\vspace{0cm} \raggedright DNN-based detector could be vulnerable to adversarial attacks. }}\\
         \hline
         \makecell{BDD strengthened using \\ Physics-based MTD \cite{LakshminarayanaCostBenefit, Lakshminarayana2021, xu2022blending, liu2020optimal} }& \checkmark & \checkmark & \makecell{Effective with \\ Large Reactance \\ Perturbations} & \makecell{Computational resources (BDD) \\ + Large periodic increase in OPF cost} & \makecell[l]{\parbox{7cm}{\vspace{0.02cm} \raggedright Applying only physics-based MTD requires sufficiently large reactance perturbations to achieve the defense goal, which lead to large increase in OPF cost.  }}\\
         \hline
         \makecell{ Static Adversarial Defense \\ Techniques \cite{li2023towards, kurakin2016adversarial, xie2017adversarial, buckman2018thermometer} } & \checkmark & \checkmark & \makecell[l]{~\checkmark~ Static Attacks \\ ~\ding{55}~ Adaptive Attacks} & \makecell{ Computational resources \\ (single DNN + adversarial training) }&\makecell[l]{ \parbox{7cm}{\raggedright Static defense could be vulnerable to adaptive attackers (e.g., CW attacks could bypass an adversarially trained model).}}\\
         \hline
         %MTD-strengthened DNN \cite{fMTD2019, sengupta2019mtdeep, qian2022ei, Morphence2021}
          \makecell{\textbf{Proposed Approach~1:} \\ \textbf{MTD-strengthened DNN} } & \checkmark & \checkmark & \makecell{Suboptimal Defense \\ (Accuracy $\leq0.942$)} & \makecell{ Computational resources \\ (multiple DNN models)} &\makecell[l]{\parbox{7cm}{\vspace{0.02cm} \raggedright Direct application of MTD-strengthened DNNs cannot achieve the desired detection accuracy and leads to high computational costs.}}\\
         \hline
        \makecell{\textbf{Proposed Approach~2:} \\ \textbf{Physics-based MTD} \\ + \textbf{MTD-strengthened DNN} } & \checkmark & \checkmark & \checkmark~(Accuracy $>0.99$) & \makecell{ Computational resources \\ (multiple DNN models) \\ + Increase in OPF cost } & \makecell[l]{\parbox{7cm}{\vspace{0.02cm} \raggedright This approach achieves high detection accuracy with manageable computational resources and minimal OPF cost increase.}}\\
         \hline
    \end{tabular} }
    \label{tbl:contributions}
    \vspace{-0.5cm}
\end{table*}

\subsection{Contributions and Paper Outline}
In this work, for the first time, we develop an MTD approach to strengthen the DNN-based attack detectors in power grids that can detect BDD and DNN-bypassing adversarial attacks. The proposed approach is inspired by similar works on defending against adversarial attacks in the context of image processing \cite{fMTD2019, sengupta2019mtdeep, qian2022ei, Morphence2021, song2021deepmtd} or malware detection \cite{rashid2023stratdef}. Specifically, we develop MTD-strengthened DNN, which deploys multiple DNN models, referred to as \emph{model pool}, instead of a single static DNN model that collaboratively makes classification decisions to detect adversarial attacks. This model pool is designed to maintain performance on clean datasets while presenting diverse decision landscapes toward adversarial attacks. The diversity among models makes it challenging for an adversarial example to bypass all DNN models simultaneously, as its transferability across different models is limited. The models are continuously updated to increase the difficulty for attackers in obtaining real-time knowledge of the models. However, \cite{Rashid2025} highlights that the model pool remains imperfect and susceptible to certain threats. Therefore, we go beyond the direct application of the MTD approach proposed in \cite{fMTD2019, sengupta2019mtdeep, qian2022ei, Morphence2021, song2021deepmtd, rashid2023stratdef} and aim to leverage the \emph{domain-specific} aspects of power grids in attack detection. To this end, we combine the design of MTD-strengthened DNN with physics-based MTD in order to enhance the attack detection effectiveness and reduce the computational costs associated with the creation of the MTD model pool. The proposed design achieves a balance between the computational costs associated with MTD-strengthened DNN and the operational costs associated with physics-based MTD. The key contributions of this work can be summarized as follows.

\begin{itemize}
    % \item We verify the threat of adversarial FDIAs in the power system. While prior works \cite{AFDIA2021, li2023towards} are limited to the DC power flow model, our work extends the analysis to AC conditions. Additionally, we study how the magnitude of BDD-bypassing FDIAs, which attackers aim to hide, influence the threat posed by developed adversarial FDIAs. 

    % We address the challenge that existing MTD-strengthened DNN approaches, especially techniques like data augmentation and retraining from strategically selected initialization, cannot be directly applied or are less efficient in strengthening DNN-based detectors in the power grid. 

    \item Developing MTD-strengthened DNN approach that detects adversarial FDI attacks against power system state estimation. By using different datasets to train the MTD model pool, we aim to reduce the transferability of adversarial attacks across the different DNNs (deployed within the model pool), thus increasing the probability of attack detection.

    \item Combining the MTD-strengthened DNN with physics-based MTD to increase the effectiveness of attack detection and reduce the computational costs associated with developing the model pool. Additionally, we discuss fast retraining approaches that enable DNNs to effectively adapt to topology reconfigurations caused by physics-based MTD. 

    \item Validating the proposed MTD approaches by performing extensive simulations on IEEE test bus systems and testing with various adversarial FDIA settings, such as those aiming to hide different magnitudes of BDD-bypassing FDIAs.

    %\item We propose generating diverse models by allocating different strengths of attack samples in each model's training dataset. These diverse models form a model pool that cooperatively detects threats, achieving feasible MTD-strengthened DNN in defending against adversarial FDIAs. Additionally, we investigate the performance of MTD-strengthened DNN using insightful metrics like transferability rate and explore the influence of incorporating adversarial training. 
    %\item We validate the effectiveness of combining DNNs with physics-based MTD in defending against adversarial FDIAs. We utilize the smallest principal angle (SPA) \cite{LakshminarayanaCostBenefit} to quantify the defense efficiency and discuss its associated operation cost. Additionally, we discuss fast retraining approaches that enable DNNs to effectively adapt to topology reconfigurations caused by physics-based MTD. 
    % \item To balance defense effectiveness and operational cost, we propose the combination of MTD-strengthened DNN and physics-based MTD. Simulation results show that combining these two techniques can achieve satisfactory defense performance with less adversarial training and a smaller SPA, which achieves a cost-benefit balance for the MTD.  
\end{itemize}

The remainder of the paper is organized as follows. Section \ref{sec: Related Work} introduces the related work. Section \ref{Preliminaries} introduces the relevant preliminaries. Section \ref{Moving Target defense against Adversarial FDIA} introduces the MTD-strengthened DNN strategies against adversarial FDIAs. Section~\ref{Physics-based MTD} introduces the combination of DNN and physics-based MTD. Section \ref{Simulation Results} presents the simulation settings and results. Section \ref{Conclusion} concludes the paper.

\vspace{-3mm}
\section{Related Work} \label{sec: Related Work}
In this section, we provide a brief survey of related works in power grid literature. Table~\ref{tbl:contributions} summarizes the novelty of our work with respect to the existing literature.

\vspace{-3mm}
\subsection{Machine Learning for FDI Attack Detection and Adversarial Attacks Against Power Grids}
Reference 
\cite{ozay2015machine} was the first to employ ML approaches to detect FDI attacks in smart grids, including perceptron, k-nearest neighbour, support vector machines (SVM), and sparse logistic regression. In \cite{he2017real}, a supervised deep learning (DL) method, namely the conditional deep belief network (CDBN), was applied for real-time FDIA detection. 
The availability of labelled datasets (especially those from the attack class) is a challenge when applying the ML approach to cybersecurity studies. 
In \cite{ZhangSemi}, researchers utilized semi-supervised DL to identify the presence of BDD-bypassing FDIAs, which requires only a few labelled measurement data in addition to unlabeled data for training. In \cite{ZhangML2020}, unsupervised DL was employed to detect cyber attacks in transactive energy systems (TES) using a deep stacked autoencoder.
To ensure the privacy of the underlying dataset, \cite{tran2023efficient} proposed a cross-silo federated learning scheme for detecting FDIAs that uses double-layer encryption and parallel computing. Furthermore, in addition to detecting the existence of FDIAs, reference \cite{wang2020locational} applied a convolutional neural network (CNN) as a multi-label classifier to identify the location of FDIAs. 

Recent works have explored the vulnerability of ML-based detectors to adversarial attacks. 
The researchers in \cite{huang2022adversarial} applied the Fast Gradient Sign Method (FGSM) and the Basic Iterative Method (BIM) methods to generate adversarial FDIAs which can bypass only the DNN-based detectors. \cite{li2023towards} considered bypassing the joint detectors and generated adversarial FDIAs using Projected Gradient Descent (PGD), which projects the adversarial perturbation into the solution space of the topology Jacobian matrix during each iterative step to bypass the BDD. Furthermore, the researchers in \cite{AFDIA2021, Tian2022target} generated adversarial FDIAs using the Carlini \& Wagner (CW) approach \cite{carlini2017towards}, which can bypass both DNN-based detectors and BDD, while also minimizing the magnitude of the adversarial perturbation to ensure that the objectives of the original BDD-bypassing FDIA remain intact. To defend against adversarial attacks, \cite{GuihaiAdv2021} applies adversarial training to strengthen the detection model in the context of smart grid demand response. Researchers in \cite{li2023towards} propose an adversarial-resilient DNN detection framework that incorporates random input padding to prevent attackers from successfully launching adversarial FDIAs. However, as noted before, these static defense mechanisms can be bypassed by sophisticated adversaries. 

\vspace{-3mm}
\subsection{MTD in Power Grids}
The topic of MTD in power grids has primarily focussed on physics-based MTD with reactance perturbation schemes being the main implementation strategy. MTD design strategy involves balancing MTD's effectiveness (i.e., ability to detect attacks), cost (i.e., the effect of MTD perturbations on the grid's operation), and its hiddenness (i.e., ensuring that the attacker cannot detect the activation of MTD) \cite{lakshminarayana2024survey}.
Metrics such as the smallest principal angle (SPA) and the rank of composite matrices are used to quantify MTD's effectiveness \cite{LakshminarayanaCostBenefit, LiuMTD2018}. The operational costs of MTD are typically quantified through increases in OPF cost or the power losses \cite{LakshminarayanaCostBenefit}, while MTD hiddenness is assessed using branch power flow consistency \cite{liu2022}. Moreover, strategies for deploying D-FACTS devices, including spanning tree methods and heuristic algorithms, are designed to enhance effectiveness while minimizing the number of D-FACTS devices required \cite{liu2020optimal, ZhangMTD2020}. Advanced models, such as adaptations for AC power flow, microgrid configurations, and game-theoretic approaches, further enhance MTD performance in practical applications \cite{CuiHidden2021, LiuMCMTD2021, LakshGT2021}.
The combination of physics-based MTD and DNN-based detection has recently been considered for power system applications.
In these works, DNN has been used as an additional tool to support MTD, such as
 for attack localization \cite{chen2022localization} or to design event-triggered MTD \cite{xu2022blending}. However, none of these works consider the vulnerability of DNNs themselves and defending against adversarial FDIAs that bypass both the BDD and DNN-based detection.

% Early studies in this area employed random reactance perturbations to obscure the attack surface but did not offer performance guarantees \cite{RahmanMTD2014}. 

% Despite these advancements, achieving complete MTD remains constrained by practical limitations, such as grid topology and meter density. Consequently, most implementations focus on incomplete but optimized MTD solutions. 

%. For example, reference\cite{chen2022localization} uses physics-based MTD to expose stealthy coordinated cyber-physical attacks (CCPAs) and employs DNNs to locate these attacks. Another study \cite{xu2022blending} introduced an event-triggered MTD method, where DNN acts as a primary detector and triggers MTD upon suspected attacks, reducing false alarms and operational frequency. Different from the current work, the primary focus of \cite{chen2022localization, xu2022blending} is on detecting BDD-bypassing FDIAs only and DNNs are used as additional tools to support attack detection. 

\vspace{-3mm}
\section{Preliminaries} \label{Preliminaries}
\subsection{Power System State Estimation} \label{Power system state estimation}

We consider a power grid consisting of a set $ \mathcal{N} = 1,2, \ldots, N$ of buses and a set $ \mathcal{L} = 1,2, \ldots, L$ of transmission lines. The power system state estimation (PSSE) finds the best estimation of the system state from the noisy measurements. 

In AC power flow model, the relationship between measurements and state variables can be represented as:
\begin{align}
     \zv = h(\sv) + \ev,
\end{align}
where $ \zv = (z_1, z_2, \ldots, z_M)$ denotes the available measurements, and $M$ is the total number of meters. In this case, the measurement $\zv \in \RR^M$ corresponds to nodal voltage magnitude, active and reactive power flow, active and reactive power injections, i.e., $\zv = [\tilde{\Vm},\tilde{\Pm_f},\tilde{\Qm_f},\tilde{\Pm},\tilde{\Qm}]^T$. The measurement error (noise) is denoted by $ \ev = (e_1, e_2, \ldots, e_M)$ which is assumed to be Gaussian. The system state consists of nodal voltage magnitudes and phase angles, i.e., $\sv = [\Vm, \thetav]$, and $h(\cdot)$ is a function vector that establishes dependencies between measured values and state variables. 

The phase angle difference is denoted as $\theta_{i,j} = \theta_i-\theta_j$, and $\Ym = \Gm + j \Bm$ denote the bus admittance matrix, where $\Gm$ and $\Bm$ denote conductance and susceptance matrices respectively. According to the observed measurements, the state variables are determined from the following weighted least square optimization problem:
\begin{equation}
    \min_{\sv} J(\sv) = (\zv-h(\sv))^T \cdot \Wm \cdot (\zv-h(\sv)).
    \label{eqn:AC WLS}
\end{equation}
The estimated state vector is $\hat{\sv} = \arg \min\limits_{\sv} J(\sv)$ and the solution $\hat{\sv}$ satisfies $\frac{\partial J(\hat{\sv})}{\partial \sv} = -2\Hm_{ac}^T(\hat{\sv})\Wm(h(\hat{\sv})-\zv) = 0$, where $\Hm_{ac}(\hat{\sv}) = \frac{\partial h(\sv)}{\partial \sv} \Big \vert_{\sv=\hat{\sv}}$ is the Jacobian matrix from the function vector $h(\sv)$. $\Wm = \diag(\sigma_1^{-2}, \sigma_2^{-2}, \ldots, \sigma_M^{-2})$ is a diagonal matrix, and $\sigma_i, i = 1, \ldots,M$ is the standard deviation of sensor measurement noise. The result is a nonlinear equation system which can be solved using an iterative process. 

In DC power flow model, the relationship between measurements and state variables can be represented as:
\begin{align}
     \zv = \Hm \thetav + \ev.
\end{align}
In this case, the measurement $\zv \in \RR^M$ consists of active power flow, reverse active power flow and active power injection, i.e. ${\zv} = [\tilde{\Pm_f},-\tilde{\Pm_f},\tilde{\Pm}]^T$, where $\Pm_f = (P_{f_b^{(1)}}, P_{f_b^{(2)}}, \ldots, P_{f_L})$, $\Pm = (P_1, P_2, \ldots, P_N)$. The state of the system is given by the voltage phase angles $\thetav = (\theta_1, \theta_2, \ldots, \theta_N)^T$. We let $l = \{ i, j\}, \ i \neq j$ denote a transmission line $l \in \mathcal{L}$ that connects bus $i$ and bus $j$, and its reactance by $x_l$, thus $P_{f_l} = \frac{1}{x_l}(\theta_i - \theta_j)$. Let $\Am \in \RR^{(N-1) \times L}$ denote the reduced branch-bus incidence matrix obtained by removing the row of the slack bus and $\Dm \in \RR^{L \times L}$ as a diagonal matrix of the reciprocals of link reactances. Then, the system's Jacobian matrix $\Hm \in \RR^{M \times N}$ is given by $\Hm =  [\Dm \Am^T;-\Dm \Am^T;\Am \Dm \Am^T]$. Using the minimum mean squared estimation method, the estimate of the system state is given by $\hat{\thetav} = (\Hm^T \Wm \Hm)^{-1} \Hm^T \Wm \zv.$ Bad data detection (BDD) compares the measurement residual, which is defined as $r =  ||\zv - h(\hat{\sv})||_2$ under AC condition and $r =  ||\zv - \Hm \hat{\thetav}||_2$ under DC condition, against a predefined threshold $\tau$ and raise alarm if $r\geq\tau$. The detection threshold $\tau$ is determined by the system operator to ensure a certain false positive rate, which is usually a small value.  

\vspace{-3mm}
\subsection{BDD-bypassing FDIA} \label{False Data Injection Attack}

A False Data Injection Attack (FDIA) injects malicious data into the measurements, misleading PSSE to obtain incorrect system states. 
We denote an FDIA vector by $\av = (a_1, a_2, \ldots, a_M)^T$. Then the compromised measurement is given by $\zv_a = {\bf z} + {\bf a}$. An attack is referred to as a BDD-bypassing attack if the residual corresponding to $\zv_a$ is no greater than the preset threshold $\tau$. Under DC condition, FDIAs of the form $\av = \Hm \cv $ are undetectable, and the estimated system state (using the under attack measurements $\zv_a$ becomes $\hat{\thetav}_c = \hat{\thetav} + \cv$, where $\cv = (c_1, c_2, \ldots, c_N)^T$ is the estimation error due to the attack. Under AC condition, an attacker can craft an undetectable FDIA as $\av = h(\hat{\sv}+\cv)-h(\hat{\sv})$.
Throughout this work, we refer to the attack vector $\av$ as \emph{BDD-bypassing FIDA}.

\vspace{-3mm}
\subsection{Physics-Based Moving Target Defense} \label{sec: Physics-Based Moving Target Defense}
Physics-based MTD is a dynamic defense strategy that changes the transmission line reactance using D-FACTS devices to invalidate the attacker's acquired knowledge to launch stealthy attacks \cite{liu2020optimal, LakshDataDrive2021,  LakshminarayanaCostBenefit}. The design of physics-based MTD consists of two phases -- (i) D-FACTS device placement, and (ii) D-FACTS device operation. The D-FACTS device placement is an offline process, which is determined using a graph theoretic approach \cite{liu2020optimal}.
Let us denote the set of transmission lines by $\mathcal{L}_D \subseteq \mathcal{L}$ on which D-FACTS devices are deployed. The selection of $\mathcal{L}_D$ can be determined using graph theory, where the target system is represented as an undirected weighted graph, with the weight of each edge determined by the linear sensitivity of transmission loss to line reactance. Deploying D-FACTS devices on the minimum-weight spanning tree of this graph minimizes the number of D-FACTS devices and optimizes the MTD effectiveness \cite{liu2020optimal}.

MTD operates by changing the transmission line reactance on $\mathcal{L}_D$, which in turn alters the system's Jacobian matrix. The D-FACTS operation is an online process that determines the level of perturbation applied in the installed D-FACTS devices. In this phase, the reactance perturbation levels are determined through an optimization formulation that minimizes operational costs while maintaining a specific level of effectiveness.
%The Smallest Principal Angle (SPA) quantifies the relationship between measurement matrices before and after MTD, with higher values indicating greater MTD effectiveness. While rank of the composite matrix also assesses MTD effectiveness, SPA is more robust in noisy environments, whereas rank-based metrics are more effective in noiseless conditions but less reliable with noise \cite{xu2021robust}. We use SPA as the effectiveness metric because it is better suited for real-world applications.
There are two main approaches to develop the MTD models -- (i) MTD designed to increase the smallest principle angle (SPA) between the column spaces of the pre- and post-perturbation measurement matrices, and (ii) MTD designed to increase the rank of the composite matrix (formed by the pre- and post-perturbation measurement matrices) \cite{lakshminarayana2024survey}. Comparing SPA and rank-based metrics, \cite{xu2021robust} found that SPA provides robust performance in noisy environments, while rank-based MTD is more effective in noiseless scenarios but less reliable with noise. The results in \cite{xu2021robust} show that, in noisy environments, the SPA method can achieve much higher accuracy against worst-case attacks and outperforms by $10\%-45\%$ against random attacks compared to the rank-based method. Since our simulations consider noisy measurement data that are reflective of real-world settings, we mainly use SPA as the effectiveness metric.

MTD perturbations, however, incur non-zero operational costs. Note that MTD utilizes pre-existing devices, making capital costs, such as D-FACTS deployment, maintenance, and upgrade expenses, negligible compared to operational costs. Additionally, these costs are device-specific and lack generic models suitable for research studies; therefore, we do not consider them in this work. As a result, operational cost is the most relevant factor for MTD implementation. In the absence of MTD, the line reactance values are set to minimize the OPF cost (and/or minimize the system power losses). Reactance perturbations due to the MTD that are designed to invalidate the attacker's knowledge will lead the system to operate away from the optimal setting, thus incurring a non-negative cost. Therefore, the MTD costs are characterized by the increase in the OPF cost due to the line perturbations. As shown in \cite{LakshminarayanaCostBenefit}, there exists a trade-off between the effectiveness of MTD's attack detection and the associated implementation costs. In general, MTD reactance perturbations that are more effective in terms of attack detection capabilities incur higher operational costs.

\vspace{-3mm}
\subsection{DNN based FDIA detection} \label{DNN based attack detection}
% \sub{Put a ref where supervised detector has been developed. \checkmark}
% Recently, there has been a surge of interest in applying DNN-based approach to detect the BDD-bypassing attacks \cite{james2018online}. 
We consider a simple setup in which the FDI attack detection is modeled as a supervised binary classification problem\footnote{The developed MTD framework can be extended to other ML-based approaches as well.}, which takes the measurements as inputs and provides a binary output -- no attack (i.e., label `0') or under attack (i.e., label `1'). Let $y = f(\zv,\omegav)$ denote a parametric function, that takes the system measurements $\zv \in \mathbb{R}^m$ as inputs and outputs a label $y \in \{0, 1\}$. Herein, $\omegav$ denotes the parameters of the DNN. Further, let $\mathcal{T} = \{ \zv^{(n)}, y^{(n)} \}^{|\mathcal{T}|}_{n = 1}$ denote the input-output pair of the training dataset, $|\mathcal{T}|$ denotes the number of training samples and subscript $n$ denotes the training sample's index. The DNN parameters are trained to minimise the cross-entropy loss function given by
\begin{align}
    J_{\mathcal{T}} (\omegav) = -\frac{1}{|\mathcal{T}|} \sum_{n = 1}^{|\mathcal{T}|} & ( y^{(n)} log(f(\zv^{(n)},\omegav))   \nonumber \\
    & + (1-y^{(n)})log(1-f(\zv^{(n)},\omegav))). \label{eqn:loss_ML}
\end{align}
The DNN model is trained offline, and the developed model is then deployed online to detect the FDIAs. 
% The DNN-based detector can effectively detect the BDD-bypassing FDIAs \cite{james2018online}.

\vspace{-3mm}
\subsection{Adversarial Attack Bypassing the BDD and DNN-Based Detection} \label{Adversarial FDIA in power system}
The focus of this paper is on adversarial FDIAs that can bypass both the BDD and the DNN-based detection. We primarily focus on white-box attacks, in which the attacker is assumed to have full knowledge of the deployed DNN models \cite{AFDIA2021}. This setting, commonly addressed in previous literature, helps system operators study the worst-case scenarios. Let $\deltav$ represent an adversarial perturbation added to $\zv_a$ such that the overall attack $\zv_{adv} = \zv + \av + \deltav$ bypasses both the BDD and the DNN-based detector. The adversarial FDIA that achieves this objective can be computed by solving the following optimization problem \cite{carlini2017towards}:
\begin{align}
\min_{\deltav} \quad & ||\deltav||_2 \label{eqn:min distence metric}\\
s.t. \quad & f(\zv_a + \deltav) = 0, \label{eqn:constrain1}\\
& f(\zv_{a}) = 1. \label{eqn:constrain2}
\end{align}
{For a fixed input $\zv_a$, the objective function \eqref{eqn:min distence metric} finds an adversarial perturbation $\deltav$ with minimum norm that misleads the target model $f$ to make an incorrect decision (i.e., mislead the DNN to associate a label 0 with $\zv_{adv}$ - constraint \eqref{eqn:constrain1}). Constraint \eqref{eqn:constrain2} implies that the DNN correctly identifies $\zv_a$ (i.e., measurements without the adversarial perturbation) as under attack.} 

In this work, we apply the solution of CW approach \cite{AFDIA2021, carlini2017towards} to solve the optimization problem in \eqref{eqn:min distence metric}, \eqref{eqn:constrain1}, \eqref{eqn:constrain2}. 
Let us denote the decision function of DNN as $f(\cdot) = \sigma(\rho(\cdot))$, where $\sigma(\cdot)$ is the softmax function employed at the DNN's output layer that assigns the labels (0 and 1), and $\rho(\cdot)$ is the output of the rest of DNN layers. 
Under the CW approach, first, the constraints  \eqref{eqn:constrain1} and \eqref{eqn:constrain2} are replaced using the following constraint:
\begin{align}
g(\zv_a + \deltav) \defines \max(\rho(\zv_a + \deltav)_1-\rho(\zv_a + \deltav)_0, 0) \leq 0,\label{eqn:g}
\end{align}
where $\rho(\zv_a + \deltav)_i$ denotes the logit of $\zv_a + \deltav$ activated for the $i$-th class (in this case, 1 denotes an ``attacked'' sample, and 0 denotes a ``normal'' sample). An adversarial measurement evades the DNN detection if it activates the $0$-th class. This occurs when $\rho(\zv_a+\deltav)_0 \geq \rho(\zv_a+\deltav)_1$ and $g(\zv_a+\deltav) \leq 0$. Then, the constraint $g$ is integrated into the optimization \eqref{eqn:min distence metric}, and the adversarial FDIA model is formulated as:
\begin{align}
    & \min_{\deltav} ||\deltav||_2 + \lambda g(\zv_a+\deltav), 
   \label{eqn:CW1}
\end{align}
where $\lambda > 0$ is a trade-off parameter to balance the magnitude of $\deltav$ and the chance to achieve $g(\zv_a+\deltav) \leq 0$.

Note that the formulation above bypasses the DNN-based detection but does not ensure bypassing the BDD. In order to ensure that the attack bypasses both the detectors, we constrain $\deltav$ as 
$\deltav = \Hm [\Id_c \odot \deltav_c],$ where $\deltav_c$ is the perturbation on the state variables, and $\Id_c \in \RR^{1 \times N}$ vector denoting the access/sparsity constraint for the adversarial perturbation given by $\Id_{c,i} = 1 \ \text{if} \ \cv_i \neq 0$ and $\Id_{c,i} = 0$ otherwise. Effectively, 
the above formulation restricts the adversarial attack to manipulate only those sensors that were accessed by the attacker to construct the BDD-bypassing attack \cite{AFDIA2021}.

\begin{algorithm}[b] 
	\caption{Adversarial FDIA} \label{alg:Adversarial attack in FDIA}
    \textbf{Input:} $\zv_{a}, f$
    \textbf{Output:} $\zv_{adv}$\\
	\begin{algorithmic}[1]
    \vspace{-0.45cm}
	\STATE Initialize $\underline \lambda, \bar \lambda, \lambda_0, \alpha, D_{min}$
	\FOR{$bs = 1:\overline{bs}$}
             \FOR{$itr = 1:\overline{itr}$}
		      \STATE $\deltav_c \leftarrow \deltav_c - \alpha  \frac{\boldsymbol{\psi}}{\deltav_c}, \zv_{adv}^\prime = \zv_a + \deltav$
                \STATE \textbf{if} {$ g(\zv_{adv}^\prime) \leq 0 \ and \ ||\bf{I}_c \odot \deltav_c||_2 \leq D_{min} $} 
                \STATE \textbf{then} $\zv_{adv} \leftarrow \zv_{adv}^\prime, D_{min} \leftarrow  ||\bf{I}_c \odot \deltav_c||_2$ \textbf{end if}
            \ENDFOR
            \STATE \textbf{if} $g(\zv_{adv}) \leq 0$ \textbf{then} $\bar \lambda \leftarrow \lambda$ \textbf{else} $\underline \lambda  \leftarrow \lambda$ \textbf{end if}
            \STATE $\lambda = (\underline \lambda + \bar \lambda)/2$
    \ENDFOR
	\STATE Return $\zv_{adv}$
	\end{algorithmic}
\end{algorithm}

\begin{comment}
    \begin{algorithm}[b] 
	\caption{Adversarial FDIA} \label{alg:Adversarial attack in FDIA}
    \textbf{Input:} $\zv_{a}, f$\\
    \textbf{Output:} $\zv_{adv}$\\
	\begin{algorithmic}[1]
    \vspace{-0.4cm}
	\STATE Initialize $\underline \lambda, \bar \lambda, \lambda_0, \alpha, D_{min}$
	\FOR{$bs = 1:\overline{bs}$}
             \FOR{$itr = 1:\overline{itr}$}
		      \STATE $\deltav_c \leftarrow \deltav_c - \alpha  \frac{\boldsymbol{\psi}}{\deltav_c}, \zv_{adv}^\prime = \zv_a + \deltav$
                \IF{$ g(\zv_{adv}^\prime) \leq 0 \ and \ ||\bf{I}_c \odot \deltav_c||_2 \leq D_{min} $} 
                   \STATE $\zv_{adv} \leftarrow \zv_{adv}^\prime, D_{min} \leftarrow  ||\bf{I}_c \odot \deltav_c||_2$
                \ENDIF
            \ENDFOR
            \IF{$g(\zv_{adv}) \leq 0$} 
            \STATE $\bar \lambda \leftarrow \lambda$   
            \ELSE 
            \STATE $\underline \lambda  \leftarrow \lambda$
            \ENDIF
            \STATE $\lambda = (\underline \lambda + \bar \lambda)/2$
    \ENDFOR
	\STATE Return $\zv_{adv}$
	\end{algorithmic}
\end{algorithm}
\end{comment}

Under AC conditions, for sufficiently small perturbations $\deltav_c$, we can make the following approximation:
\begin{align}
   h(\hat{\sv}_a + \Id_c \odot \deltav_c) &\approx h(\hat{\sv}_a) + \frac{\partial h(\hat{\sv}_a)}{\partial \sv_a} [\Id_c \odot \deltav_c] \nonumber \\
   &= h(\hat{\sv}_a) + \Hm_{ac}(\hat{\sv}_a)[\Id_c \odot \deltav_c],
   \label{eqn:S-AFDIA-AC}
\end{align}
where $\hat{\sv}_a = \hat{\sv} + \cv$. Consequently, the measurement residuals will not increase if $\deltav = \Hm_{ac}(\hat{\sv}_a)[\bf{I}_c \odot \deltav_c]$. The integration of sparsity limitations results in the adversarial FDIA model aiming to find a feasible state perturbation $\deltav_c$ through the resolution of the subsequent optimization problem: 
\begin{align}
   \psiv = \min_{\deltav_c} ||\Id_c \odot \deltav_c||_2 + \lambda g(\zv_a+\deltav).
   \label{eqn:S-AFDIA}
\end{align}

The solution to this problem (detailed in Algorithm~\ref{alg:Adversarial attack in FDIA}) involves the application of Projected Gradient Descent (PGD), which is a gradient-based iterative solver for constrained optimization. Additionally, a binary search algorithm is employed to fine-tune the regularization parameter, ensuring a balance between attacks' effectiveness and stealthiness.

\begin{figure}[t]
	\centering
	\includegraphics[width=0.45\textwidth]{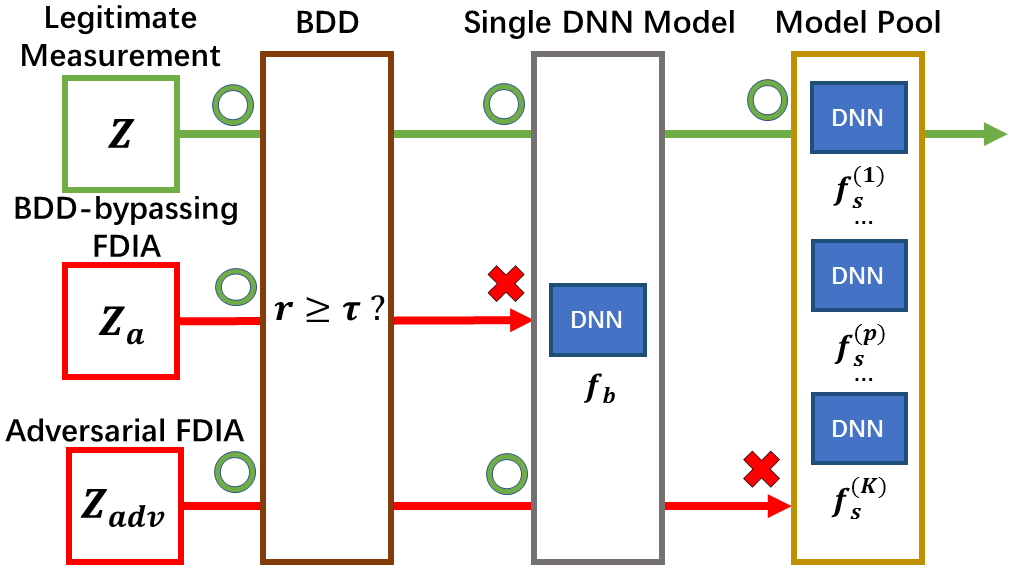}
	\caption{The framework of attack detection}
	\label{fig:framework2.png}
 \vspace{-0.7cm}
\end{figure}

\vspace{-3mm}
\section{MTD Design to Defend Against Adversarial FDIAs} \label{Moving Target defense against Adversarial FDIA}

Moving Target Defense (MTD) is a defense technique that dynamically reconfigures the system or model parameters to invalidate the knowledge that the attackers use to craft stealthy attacks. Adversarial attacks (such as those described in Section~\ref{Adversarial FDIA in power system}) are crafted by iteratively probing a \emph{fixed target} model to learn its decision function. MTD transforms the model into a \emph{moving target} by regularly altering the decision function to enhance the model's resilience against adversarial attacks.

%the decisions from multiple models are combined 
%\chen{the models in the pool apply majority vote to identify the existence of the attacks.}
% \sub{need to be precise on this.}

In our specific context, the core idea is to deploy multiple DNN models, referred to as \emph{model pool}, instead of a single static DNN model (as traditionally deployed in ML-based detection) as depicted in Figure~\ref{fig:framework2.png}.  Then, during the online inference phase, the decisions from the model pool are combined to make the final decision (details specified in Section~\ref{sec:Design of MTD against A-FDIA}). 
The deployed model pool must ensure the following criteria -- (i) they must be able to maintain accuracy on the clean examples $\zv$ (such that false alarms are minimized), (ii) ensure that adversarial examples $\zv_{adv}$ are detected with high accuracy. Furthermore, the pool of models is periodically updated so that an adversary's knowledge of the DNN parameters is invalidated. In this way, the MTD design introduces randomness to the decision boundary of the baseline DNN and generates diverse DNN models that cooperate to detect adversarial FDI attacks. In the following section, we detail the design principle of MTD-strengthened DNN against adversarial FDIAs.

\vspace{-3mm}
\subsection{Design of MTD-strengthened DNN Against Adversarial FDIAs} \label{sec:Design of MTD against A-FDIA}

\begin{figure}[t]
	\centering
    \vspace{-0.5cm}
	\includegraphics[width=0.5\textwidth]{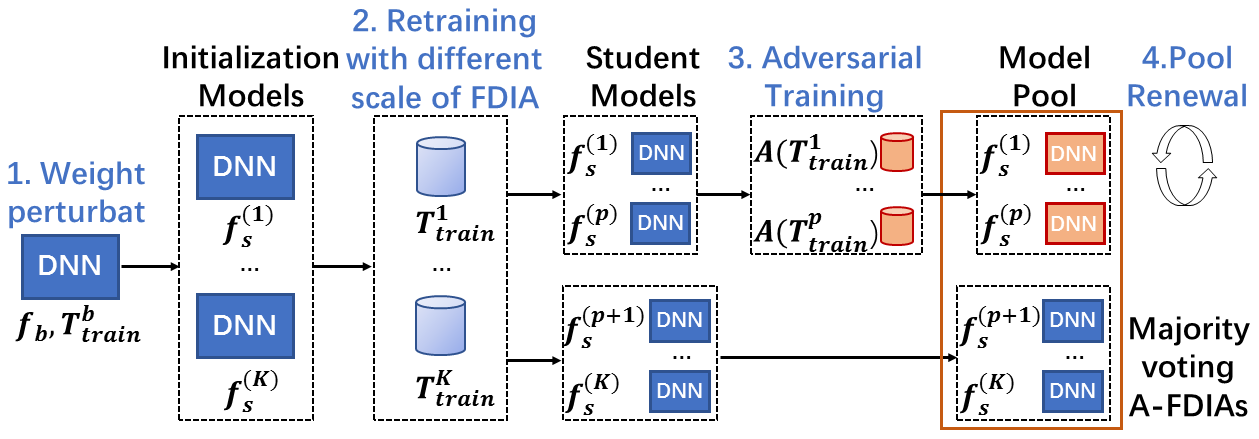}
    \vspace{-0.5cm}
	\caption{The framework of MTD-strengthened DNN}
	\label{fig:Morphence}
 \vspace{-0.5cm}
\end{figure}

The overall framework of the proposed DNN-based  MTD is illustrated in Figure~\ref{fig:Morphence} and is adapted from \cite{Morphence2021}. The process begins with the development of a base model $f_b(\omegav_b)$, representing a DNN-based detector in our specific context. This base model is trained to differentiate between legitimate measurements ($\zv$) and measurements with BDD-bypassing attacks ($\zv_a$). To this end, we used supervised machine learning to train the model parameters $\omegav_b$ using datasets. We denote the dataset used to train the base model by $\mathcal{T}_{train}^b$ (the details of the dataset generation are specified in Section~\ref{Simulation Results}).  Subsequently, several student models $f_s = \{f_s^{(1) } (\omegav_s^{(1)}), f_s^{(2)} (\omegav_s^{(2)}), \dots, f_s^{(K)} (\omegav_s^{(K)})\}$ are created from the base model, where $\omegav_s^{(k)}$ are the weights associated with the student model $k,$ and $K$ is the total number of student models deployed. These student models are derived from the base model using the following steps.

In Step 1, a random perturbation $\epsilon$ (e.g., Laplace noise) is introduced to the weights of the base model ($\omegav_b$), i.e., $\omegav_s^{(k)} = \omegav_b + \epsilon^{(k)}, \ k = 1,\dots, K$. Note that due to the random perturbations added to the weights of the base model, the classification accuracy of the student models will diminish. In order to improve accuracy, the student models are retrained in the second step. Note that due to the randomness associated with the DNN training (i.e., randomness in the initialization of the student model's weights in Step 1 and the training process, such as stochastic gradient descent), the final weights of the student models will be different from each other as well as that of the base model. Thus, any adversarial attack that bypasses the student model $i$ is unlikely to bypass the student model $j.$ In order to further ensure that the weights of the student models are different from each other, we use different datasets in the retraining process of the individual student models, where the datasets differ in the way in which the BDD-bypassing attacks are generated. We denote the training dataset used in retraining student model $k$ by  $\mathcal{T}_{train}^{(k)}.$ The details of how these different datasets are generated are specified in Section~\ref{Simulation Results}. The student models retrained on these distinct datasets exhibit sufficient diversity, reducing the transferability of adversarial examples among them. At the same time, they maintain the accuracy of identifying BDD-bypassing FDIA.

Step 3 involves applying adversarial training to enhance the robustness of this approach. In Step 2, student models with varying decision boundaries are deployed by retraining them on different magnitudes of BDD-bypassing FDIAs. However, legitimate training samples are still generated from the same distribution. The exclusive use of legitimate training samples results in similarities between the student models, making them remain susceptible to some adversarial attacks (e.g., one-step evasion attacks) \cite{Morphence2021}. Adversarial training, which is a widely adopted technique to harden models against adversarial attacks, is applied to further reduce the transferability of these attacks.
The fundamental concept involves generating and incorporating adversarial examples into the training dataset during the training process, as outlined in \cite{kurakin2016adversarial}. In the scheme of MTD, a subset of student models, denoted as $p < K$, are retrained using adversarial training. Notably, $p$ and $K$ are hyperparameters of the MTD, determined by the system operator, and their influences are investigated in Section~\ref{Simulation Results}. After that, the developed student models are integrated through a majority voting mechanism. 

%Unlike the approach in \cite{Morphence2021}, where majority voting was not employed because the votes from adversarial trained models may lead to accuracy decay on their clean queries. We have not observed apparent accuracy decay resulting from adversarial training in identifying BDD-bypassing FDIA. Consequently, majority voting would be a convenient and straightforward method for collaboration among student models.

Step 4 involves the periodic renewal of the model pool. Given sufficient time, attackers may accumulate knowledge about the current model pool, posing a risk to the proposed MTD scheme. To mitigate this threat effectively, the model pool must be updated at regular intervals. This proactive approach serves to eliminate the potential for attackers to exploit static configurations, enhancing the resilience of the overall MTD.

\vspace{-3mm}
\subsection{Combining DNN with Physics-Based MTD} \label{Physics-based MTD} 

\begin{figure}[t]
	\centering
	\includegraphics[width=0.5\textwidth]{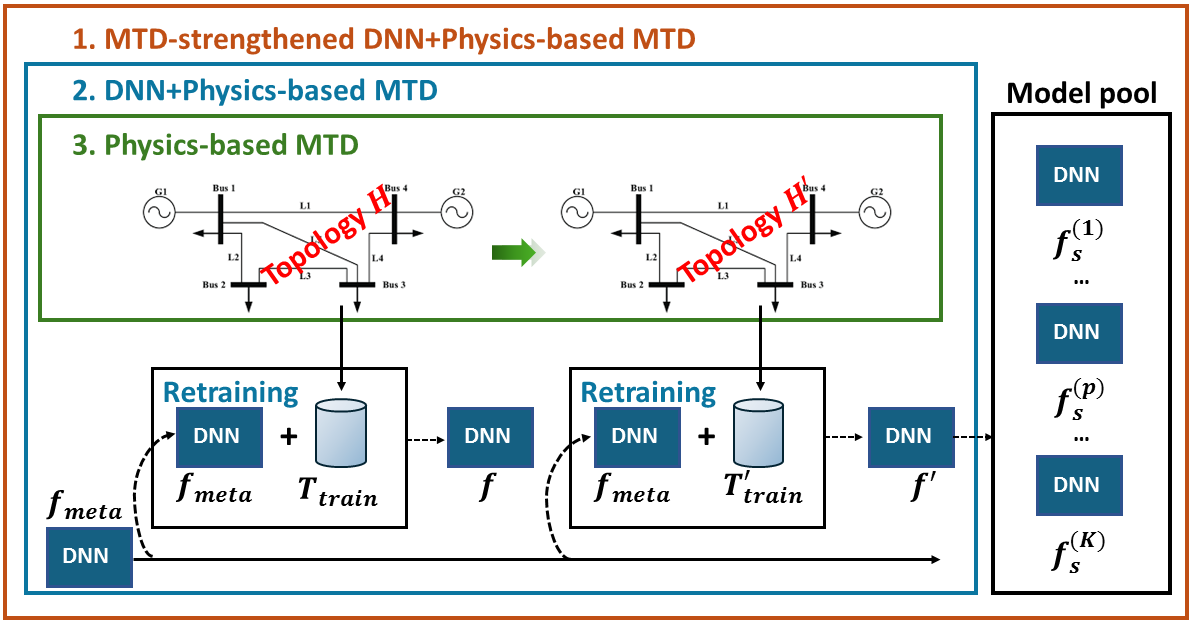}
	\caption{The framework of integrating physics-based MTD}
	\label{fig:DNN physic MTD}
 \vspace{-0.5cm}
\end{figure}

Next, we propose combining the DNN-strengthened MTD with the physics-based approach to further improve MTD's effectiveness. Furtheremore, although the MTD strategy proposed in Section~\ref{sec:Design of MTD against A-FDIA} is effective, the creation of several student models and retraining/adversarial training incurs significant computational time and memory. Combining the MTD design with physics-based MTD also significantly reduces the computational costs. 
The concept behind the physics-based MTD is described in Section~\ref{sec: Physics-Based Moving Target Defense}. While it is possible to achieve high efficiency by implementing solely the physics-based MTD approach, they also incur high operational costs (i.e., increasing the system's OPF cost). The proposed combination of DNN and physics-based MTD aims to achieve the \emph{best of both worlds} in terms of achieving high detection accuracy while keeping the operational costs low. The overall framework is shown in Figure~\ref{fig:DNN physic MTD}.

% system operator uses the new system configuration as input to a power grid simulator and generates a few new data samples for DNN retraining. 

In the combined scheme, the operator first applies reactance perturbation to implement the physics-based MTD, which changes the Jacobian matrix of the power system from $\Hm$ to $\Hm^\prime$. Following this, the base model of the DNN-based detector $f_b (\omegav_b)$ is retrained and adapted to the new system configuration. We denote the base model corresponding to the setting $\Hm^\prime$ by $f^{\prime}_b (\omegav^{\prime}_b).$ Following the adaptation of the base model, new student models are created from $f^{\prime}_b (\omegav^{\prime}_b)$ using the methodology described in the previous subsection. The attacker is assumed to have the knowledge of system parameters corresponding to $\Hm$ and $f_b (\omegav_b)$. Note that changing the system topology and the adaptation of the DNN base model can invalidate the attacker's knowledge. Developing student models further from $f^{\prime}_b (\omegav^{\prime}_b)$ strengthens the defense further. 
Consequently, adversarial attacks designed based on this information are less likely to remain effective in bypassing the new models. Notably, our simulations reveal that we require a significantly reduced number of student models and adversarial training to achieve high detection rates as compared to the original MTD-strengthened DNN design in Section~\ref{sec:Design of MTD against A-FDIA}.

 % under the current system topology while presenting different decision boundaries and adversarial attack surfaces compared to models developed based on the previous system topology (before triggering physics-based MTD). 

% \sub{This is really weak motivation; every solution must have a need.\checkmark}
%Other than applying MTD technique to strengthen DNN, as we detailed in Section~\ref{sec:Design of MTD against A-FDIA}, DNN can combine with physics-based MTD to defend against adversarial attacks. 

Note that the adaptation of the base model from $f_b (\omegav_b)$ to $f^{\prime}_b (\omegav^{\prime}_b)$ itself incurs computational costs. To minimize the overhead, we propose the application of meta-learning to accelerate the DNN retraining process, which enables rapid adaptation to the new configuration using a small number of training samples during the retaining \cite{FinnMAML2017}.  Specifically, meta-learning is a training methodology suited for learning a series of related tasks. Developing base models under different topologies (with different reactance settings) can be viewed as a series of related but different learning tasks. Meta-learning has proven effective in adapting DNNs for optimal power flow (OPF) prediction following topology reconfigurations \cite{chen2022meta}. The meta-learning algorithm consists of two main phases: an offline training phase and an online training phase. During the offline training phase, a meta model $f_\text{meta}$ is generated, and its parameters are optimized to minimize a carefully designed loss function, which ensures that $f_\text{meta}$ learns internal features that are broadly applicable to all tasks at hand rather than a specific task. Then, during the online training phase, the weights of $f_\text{meta}$ serve as initialization parameters, and meta-learning leverages these initialization parameters to quickly adjust a base model's parameters (e.g., $\omegav^{\prime}_b$) to a new task (adapt to new system configuration) with only a few gradient updates and a small number of training samples. The retrained base models (e.g., $f^\prime$) can achieve good performance in identifying FDIA in their corresponding system configurations (e.g., $\Hm^\prime$). We omit the detailed algorithm description and refer the readers to reference \cite{chen2022meta}. Thus, this method is well-suited to adapt DNN-based FDIA detection under planned topology reconfigurations such as those led by physics-based MTD.

\begin{figure}[t]
    \centering
    \includegraphics[width=0.5\textwidth]{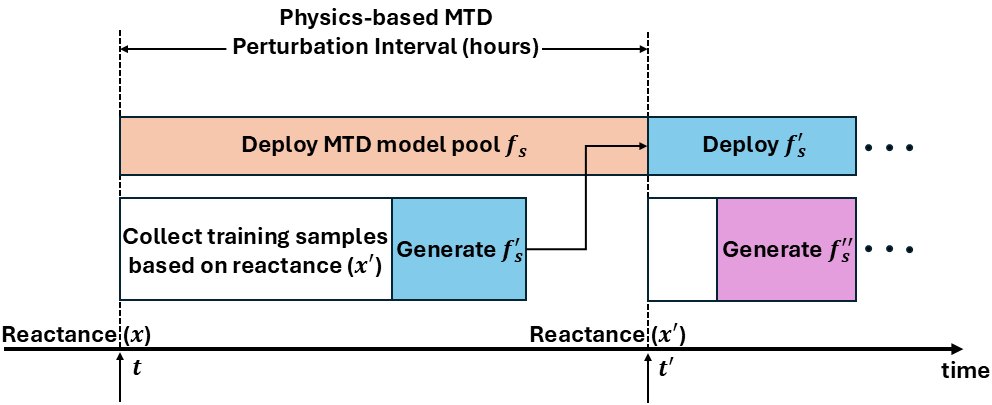}
    \caption{Timeline of overall defense}
    \label{fig: timeline of overall defense}
    \vspace{-0.5cm}
\end{figure}

The timeline of the overall defense is illustrated in Figure~\ref{fig: timeline of overall defense}. Recall that the proposed defense strategy integrates a physics-based MTD (reactance perturbations) and generates a model pool for each reactance perturbation setting. Firstly, note that the time interval between the reactance perturbation depends on the attacker's ability to learn the system parameters. Specifically, if the reactance settings are changed before the attacker can gather sufficient information to learn the new settings, then the MTD remains effective. Existing works have analysed this problem \cite{LakshGT2021}, and estimated that the time interval between reactance changes in the order of hours is sufficient to maintain MTD's effectiveness (as shown in Figure~\ref{fig: timeline of overall defense}). We now explain how the MTD model pool generation can be incorporated into this setting. Note that the physics-based MTD involves planned topology perturbations, which can be generated based on the current reactance settings and the effectiveness metric. Thus, during the reactance settings $\xv$ (interval corresponding to the orange bar), the operator can compute the new reactance settings $\xv^{\prime}$ and pre-generate a new model pool. Then, when the physics-based MTD is triggered, the previously deployed MTD model pool automatically expires and is seamlessly replaced by the newly generated model pool. To summarize, the generation of the new MTD model pool only needs to be completed before the subsequent activation of the physics-based MTD. This mechanism ensures that the proposed approach is practical and time-efficient for real-world applications.

\vspace{-3mm}
\section{Simulation Results} \label{Simulation Results}

In this section, we assess the performance of the proposed MTD strategies against adversarial FDIAs. We start by examining the effectiveness of the MTD-strengthened DNN and the impact of its hyperparameters. Then, we evaluate the effectiveness of combining the DNN with physics-based MTD and analyze the associated operational costs.
The simulations are conducted on a standard IEEE 14, 30, 118-bus system considering both DC and AC conditions. Note that while the test data used in this work are synthetic, they are specifically designed to emulate real-world conditions. We generate these datasets using the MATPOWER simulator \cite{Matpower2011}, a widely used tool in power system planning and offline analysis. We carefully configure key parameters, including measurement noise, load variation limits, and branch reactance perturbation limits. Furthermore, our attack identification approach is based on independent measurement samples and does not rely on the time-sequence information in the load profile. As a result, incorporating real-world load profiles into the test data is expected to yield similar outcomes. This is a standard experimental setting that is applied in existing references \cite{LakshminarayanaCostBenefit, xu2022blending}. Therefore, despite the synthetic nature of the test data, our testing results can effectively demonstrate the usability and applicability of our approach in real-world scenarios.

\vspace{-3mm}
\subsection{Simulation Setup} \label{Simulation setups}
 
\textbf{Dataset Generation for Normal System Operation:} We assume that loads on each bus in the test systems are uniformly distributed between 80\% and 120\% of their base (default) values and the generations are maintained at optimal dispatch to achieve optimal power flow. 
%The measurement vectors encompass voltage magnitude and phase angle 
% \sub{You mean the system state? \checkmark}. 
%The voltage magnitudes are presented using per unit (p.u.) values, which range within $[0.94, 1.06]$, and the phase angles are expressed in radians (rad), ranging within $[-\pi, \pi]$. 
Measurement error is assumed to follow a zero-mean Gaussian distribution $\ev \sim \mathcal{N}(0, 0.02)$. Assuming that each test system is fully measured, we generate legitimate measurements following the approach described in Section~\ref{Power system state estimation} and associate the label $0$.

% A BDD-bypassing attack is named as a base FDIA when it is utilized to develop adversarial perturbation. 

\textbf{Dataset Generation for BDD-bypassing FDIAs:} Subsequently, we create BDD-bypassing FDIAs according to the method detailed in Section~\ref{False Data Injection Attack}. To represent attack sparsity, the attackers are assumed to be capable of compromising up to half of the system states. 
%In the common setting, attackers are limited to distorting updates to half of the state variables . This is achieved by setting the sparsity of the state attack within the range $[1, \frac{|\sv|}{2}]$, where $|\sv|$ denotes the dimension of the state variables. 
Without the loss of generality, we assume that the measurements corresponding to the reference bus are not under attack. The magnitude of a state attack is assumed to follow a Gaussian distribution, i.e.,  $c_i \sim \mathcal{N}(0, \nu^2)$. We use the standard deviation $\nu$ to reflect the magnitude of the state attack vector ($\cv$). We generate DC attack measurements as $\zv_a = \zv + \Hm\cv$ and AC attack measurements as $\zv_a = \zv + h(\sv+\cv)-h(\sv)$ and then associate the label $1$.

%***We create different magnitudes of BDD-bypassing FDIA by modifying $\nu$ within the range $[0.05, 0.3]$.*** 

\textbf{Dataset Generation for Adversarial FDIAs:} 
After generating BDD-bypassing FDIAs and the corresponding target DNN as in Section~\ref{DNN based attack detection}, we can generate adversarial perturbations $\deltav$ following Algorithm~\ref{alg:Adversarial attack in FDIA} with its parameters setting as $\underline \lambda = 0, \bar \lambda = 100, \lambda_0 = 0.5, \alpha = 0.01, D_{min} = \infty, \overline{bs} = 5, \overline{its} = 200$. Then we have adversarial measurements $\zv_{adv} = \zv + \av + \deltav,$  and associate a label $1$. 
%Adversarial FDIAs are obtained by adding adversarial perturbations to the corresponding base FDIAs. Adversarial measurements are obtained by adding adversarial attack vectors to legitimate measurements and are labelled as $1$. 

%covering $20$ dimensions for active and reactive power flow and $14$ dimensions for voltage magnitude, active and reactive power injection

% For the 14-bus system, under DC power flow, the number of neurons in the DNN's input layer is set to $54$, which corresponds to the measurement vector ${\zv} = [\tilde{\Pm_f},-\tilde{\Pm_f},\tilde{\Pm}]^T$. Under AC power flow, the number of neurons in the DNN's input layer is adjusted to $82$, which aligns with the measurement vector $\zv =[\tilde{\Vm},\tilde{\Pm_f},\tilde{\Qm_f},\tilde{\Pm},\tilde{\Qm}]^T$.

% As a binary classifier, the output vector size is fixed at $2$, corresponding to the legitimate label $0$ and the false label $1$. 

\textbf{Implementation of DNNs:} The DNN-based detectors and MTD strategy are developed using the PyTorch framework. For developing the DNN-based detectors, we utilize a fully connected neural network, as detailed in Table~\ref{tbl:NN_arch}. The sizes of the input and output layers are customized to match the dimensions of the dataset. Under the DC power flow, the number of neurons in the DNN's input layer correspond to the dimension of the measurement vector given by ${\zv} = [\tilde{\Pm_f},-\tilde{\Pm_f},\tilde{\Pm}]^T$. Under AC power flow, this aligns with the measurement vector $\zv =[\tilde{\Vm},\tilde{\Pm_f},\tilde{\Qm_f},\tilde{\Pm},\tilde{\Qm}]^T$. The ReLU activation function is applied to the hidden layers, while the softmax activation function is employed at the output layer. 

\begin{table}[t]
	\centering
 \caption{The DNN structure used for attack detection.}
  \resizebox{0.8\columnwidth}{!}{
    	\begin{tabular}{|c|c|c|c|}
		\hline
        Test system & Input layer & hidden layers & Output layer \\
		\hline
		14-bus & 82(AC) 54(DC) & 100/50/25 & 2 \\
		\hline
        30-bus & 172(AC) 112(DC) & 200/100/50 & 2 \\ 
        \hline
        118-bus & 726(AC) 490(DC) & 800/400/100 & 2 \\
        \hline
	\end{tabular}}
	\label{tbl:NN_arch}
 \vspace{-0.5cm}
\end{table}

\vspace{-3mm}
\subsection{Evaluation Metrics}

We assess the effectiveness of ``combining DNN with physics-based MTD'' using its recall rate ($R$) on adversarial FDIAs and assess the performance of ``MTD-strengthened DNN'' using two metrics: $R$ and the transferability rate ($\eta$) of adversarial FDIAs on the model pool. The metric $\eta$ assesses how an adversarial FDIA can transfer between student models in the MTD model pool. Consider a set of adversarial FDIAs $\mathcal{T}_{\zv_{adv}} = \{ \zv_{adv}^{(n)} \}^{|\mathcal{T}_{\zv_{adv}}|}_{n = 1}$ constructed using the base DNN model $f_b$ (whose parameters can be obtained by the attackers). Let $N_{adv}(f_s^{(i)})$ denote the amount of adversarial measurements in $\mathcal{T}_{\zv_{adv}}$ that can evade the $i$-th student model and $N_{adv}(f_s^{(i)} \rightarrow f_s^{(j)})$ denote the amount of adversarial measurements in $\mathcal{T}_{\zv_{adv}}$ that can simultaneously evasive both the $i$-th and $j$-th student models. Then the transferability rate (for adversarial FDIAs in $\mathcal{T}_{\zv_{adv}}$) between $f_s^{(i)}$ and $f_s^{(j)}$ can be computed as:
$\eta_{i,j} = \frac{N_{adv}(f_s^{(i)} \rightarrow f_s^{(j)})}{N_{adv}(f_s^{(i)})}, $
and the average transferability rate among all student models (with a total number of $K$) can be computed as:
$\eta_{av} = \frac{1}{K(K-1)} \sum_{i=1}^K \sum_{\underset{j=1}{j \neq i}}^K \eta_{i,j}.$
An MTD model pool with lower $\eta_{av}$ values exhibits greater diversity among student models and overall detection effectiveness. This metric provides an insightful view of the performance of MTD-strengthened DNN.

\vspace{-3mm}
\subsection{Simulation Results} \label{sec: Simulation Results}

First, we perform simulations using the DC power flow model. We develop the MTD-strengthened DNN according to the approach detailed in Section~\ref{sec:Design of MTD against A-FDIA}. Firstly, we develop a base model $f_b$ on a training dataset, which is composed of $5000$ legitimate measurements and $5000$ BDD-bypassing FDIA measurements, where the BDD-bypassing FDIAs are constructed as in Section~\ref{Simulation setups} by setting $\nu_{f_b}=0.05$. 
Secondly, we develop $K$ student models $f_s = \{f_s^{(1) } (\omegav_s^{(1)}), f_s^{(2)} (\omegav_s^{(2)}), \dots, f_s^{(K)} (\omegav_s^{(K)}) \}$ by introducing random perturbations to $\omegav_b$, i.e., $\omegav_s^{(k)} = \omegav_b + \epsilon^{(k)}, \ k = 1,\dots, K, \ \epsilon^{(k)} \sim \mathcal{U}(-0.1\omegav_b, 0.1\omegav_b)$ (here in $\mathcal{U}$ denotes the uniform distribution). 
Thirdly, the $K$ student models are retrained using different datasets (in order to reduce the transferability). The dataset includes $5000$ legitimate measurements ($\zv$) and $5000$ BDD-bypassing measurements ($\zv_a = \zv + \av$). We construct $\av$ as in Section~\ref{Simulation setups} by choosing $\nu_{f_s^{(k)}} \sim \mathcal{U}(0.05, 0.3)$ for the $K$ student models (a different value picked for each model). Fourthly, we apply the adversarial training approach to retrain $p$ student models, whereas a standard retraining approach is used to retrain the remaining $K-p$ student models. The retrained student models form a model pool that cooperatively identifies attacks using a majority voting mechanism.

For testing, we generate four adversarial FDIA datasets: $\mathcal{T}{\zv_{adv}, f_b, \nu_1}$, $\mathcal{T}{\zv_{adv}, f_b, \nu_2}$, $\mathcal{T}{\zv_{adv}, f_b, \nu_3}$ and $\mathcal{T}{\zv_{adv}, f_b, \nu_4}$, which aim to hide the different magnitude of BDD-bypassing FDIAs, i.e., $\{\nu_1=0.05, \nu_2=0.1, \nu_3=0.2, \nu_4=0.3\}$ from base model, respectively. Each testing set contains $1000$ samples. In the following description, we refer $\nu_1, \nu_2, \nu_3, \nu_4$ to these four testing sets for simplification. 

%We use three results to illustrate the effectiveness of the proposed MTD approach. First, we plot the pairwise transferability rate $\eta_{i,j}$ by considering four of the student models in Figure~\ref{fig:confusion matrix}. We also plot the recall rate and average transferability rate $\eta$ in Figures~\ref{fig:14DCn} and Figure~\ref{fig:14DCp}. In Figure~\ref{fig:confusion matrix}, it can be observed that while adversarial FDIAs can successfully evade the targetted model, they do not effectively transfer to models other than their targeted ones. This validates our approach to constructing the model pool for MTD.

%We use two results to illustrate the effectiveness of the proposed MTD approach. We also plot the recall rate and average transferability rate $\eta$ in Figures~\ref{fig:14DCn} and Figure~\ref{fig:14DCp}. 

\begin{comment}

\begin{figure}[t]
	\centering
	\includegraphics[width=0.49\textwidth]{confusion matrix.png}
	\caption{The transferability rate between models.}
	\label{fig:confusion matrix}
 \vspace{-0.5cm}
\end{figure}

\end{comment}

\begin{figure}[t]
  \centering
  \begin{subfigure}{0.241\textwidth}
    \includegraphics[width=\linewidth]{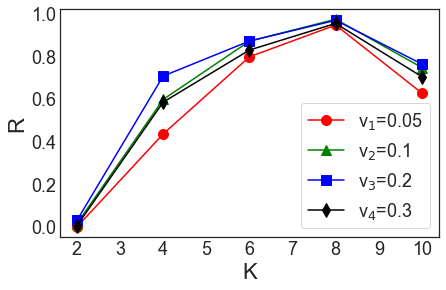}
    \vspace{-0.5cm}
    \label{fig:14DC-DR-n4}
  \end{subfigure}
  \begin{subfigure}{0.241\textwidth}
    \includegraphics[width=\linewidth]{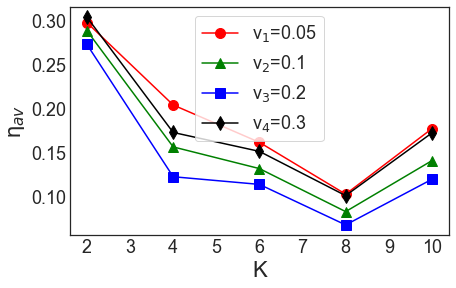}
    \vspace{-0.6cm}
    \label{fig:14DC-AT-n4}
  \end{subfigure}
  \caption{The performance of MTD-strengthened DNN over the number of student models (14-bus DC, $p=K/2$).}
  \label{fig:14DCn}
  \vspace{-0.2cm}
\end{figure}

\begin{figure}[t]
  \centering
  \begin{subfigure}{0.241\textwidth}
    \includegraphics[width=\linewidth]{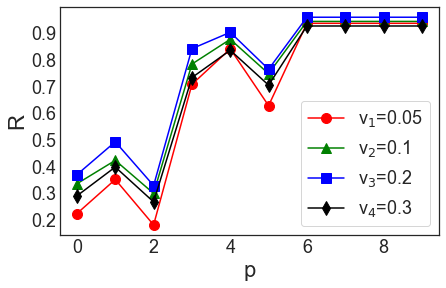}
    \vspace{-0.5cm}
    \label{fig:14DC-DR-p4}
  \end{subfigure}
  \begin{subfigure}{0.241\textwidth}
    \includegraphics[width=\linewidth]{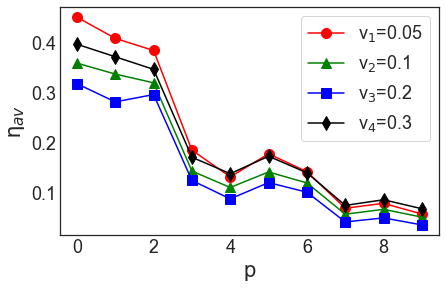}
    \vspace{-0.6cm}
    \label{fig:14DC-AT-p4}
  \end{subfigure}
  \caption{The performance of MTD-strengthened DNN over the number of adversarially trained models (14-bus DC, $K=10$)}
  \label{fig:14DCp}
  \vspace{-0.2cm}
\end{figure}

Firstly, we illustrate the effectiveness of the proposed MTD approach using two results, the recall rate and the average transferability rate $\eta$, which are plotted in Figures~\ref{fig:14DCn} and \ref{fig:14DCp}. In Figure~\ref{fig:14DCn}, the performance of MTD-strengthened DNN is plotted as a function of the number of student models $K$. In general, the MTD performance improves with an increase in the number of student models, as observed by the increasing recall rate in Figure~\ref{fig:14DCn}(a) and the decrease in average transferability rate in Figure~\ref{fig:14DCn}(b). In Figure~\ref{fig:14DCp}, the performance of MTD-strengthened DNN is depicted as a function of the proportion of adversarially trained models $p$ within the model pool. The results demonstrate an enhancement in defense effectiveness by increasing the value of $p$, with improvement plateauing when $p \geq 6$. However, increasing the value of $p$ will dramatically increase the time consumption, as shown in Table~\ref{tbl:time consumption}, resulting in a trade-off when selecting the value of $p$. Additionally, the execution times of the deployed DNN-based detection are shown in Table~\ref{tbl:execution times}. It can be observed that applying the MTD pool does not significantly increase the execution time compared to using a single DNN.

%\arrayrulecolor{blue}

\begin{comment}
    
\begin{table}[!t]
	\centering
 \caption{\chen{Time consumption for developing MTD-strengthened DNN}}
 \resizebox{\columnwidth}{!}{
    	\begin{tabular}{|c|c|c|c|c|}
		\hline
          & \multicolumn{4}{c|}{Time consumption (second)} \\
          \cline{2-5}
          &\multicolumn{2}{c|}{DC} & \multicolumn{2}{c|}{AC}\\
          \cline{2-5}
		 & $p=0, K=10$   & for $p+1$ & $p=0, K=10$   & for $p+1$ \\
		\hline
		14-bus & $1072$ &  $+460$  & $1095$ &  $+556$ \\
		\hline
		30-bus & $1119$ &  $+541$  & $2079$ &  $+667$ \\
		\hline
		118-bus & $2826$ & $+783$ & $4711$ & $+934$  \\
		\hline
	\end{tabular}
 }
\label{tbl:time consumption}
\vspace{-0.2cm}
\end{table} 

\begin{table}[!t]
	\centering
 \caption{\chen{Execution times of DNN-based detection}}
 \resizebox{\columnwidth}{!}{
    	\begin{tabular}{|c|c|c|c|c|}
		\hline
          & \multicolumn{4}{c|}{Execution Time (second)} \\
          \cline{2-5}
          &\multicolumn{2}{c|}{DC} & \multicolumn{2}{c|}{AC}\\
          \cline{2-5}
		 & Single DNN & MTD pool (p=5, K=10) & Single DNN & MTD pool (p=5, K=10)\\
		\hline
		14-bus & $1.39 \times 10^{-5} $ & $1.25 \times 10^{-4}$ & $1.34 \times 10^{-5} $ & $5.70 \times 10^{-5} $ \\
		\hline
		30-bus & $1.79 \times 10^{-5} $ & $9.73 \times 10^{-5} $ & $1.19 \times 10^{-5} $   & $8.35 \times 10^{-5} $ \\
		\hline
		118-bus & $1.89 \times 10^{-5} $ & $8.95 \times 10^{-5} $ & $1.95 \times 10^{-5} $   & $1.05 \times 10^{-4}$ \\
		\hline
	\end{tabular}
 }
\label{tbl:execution times}
\vspace{-0.2cm}
\end{table} 

\end{comment}

\begin{table}[!t]
	\centering
 \caption{Time consumption for developing MTD-strengthened DNN}
 \resizebox{0.8\columnwidth}{!}{
    	\begin{tabular}{|c|c|c|c|c|}
		\hline
          & \multicolumn{4}{c|}{Time consumption (second)} \\
          \cline{2-5}
          &\multicolumn{2}{c|}{DC} & \multicolumn{2}{c|}{AC}\\
          \cline{2-5}
		 & $p=0, K=10$   & for $p+1$ & $p=0, K=10$   & for $p+1$ \\
		\hline
		14-bus & $1072$ &  $+460$  & $1095$ &  $+556$ \\
		\hline
		30-bus & $1119$ &  $+541$  & $2079$ &  $+667$ \\
		\hline
		118-bus & $2826$ & $+783$ & $4711$ & $+934$  \\
		\hline
	\end{tabular}
 }
\label{tbl:time consumption}
\vspace{-0.2cm}
\end{table}

\begin{table}[!t]
	\centering
 \caption{Execution times of DNN-based detection}
 \resizebox{\columnwidth}{!}{
    	\begin{tabular}{|c|c|c|c|c|}
		\hline
          & \multicolumn{4}{c|}{Execution Time (second)} \\
          \cline{2-5}
          &\multicolumn{2}{c|}{DC} & \multicolumn{2}{c|}{AC}\\
          \cline{2-5}
		 & Single DNN & MTD pool (p=5, K=10) & Single DNN & MTD pool (p=5, K=10)\\
		\hline
		14-bus & $1.39 \times 10^{-5} $ & $1.25 \times 10^{-4}$ & $1.34 \times 10^{-5} $ & $5.70 \times 10^{-5} $ \\
		\hline
		30-bus & $1.79 \times 10^{-5} $ & $9.73 \times 10^{-5} $ & $1.19 \times 10^{-5} $   & $8.35 \times 10^{-5} $ \\
		\hline
		118-bus & $1.89 \times 10^{-5} $ & $8.95 \times 10^{-5} $ & $1.95 \times 10^{-5} $   & $1.05 \times 10^{-4}$ \\
		\hline
	\end{tabular}
 }
\label{tbl:execution times}
\vspace{-0.5cm}
\end{table}

\arrayrulecolor{black}

\begin{figure}[t]
  \centering
  \begin{subfigure}{0.241\textwidth}
    \includegraphics[width=\linewidth]{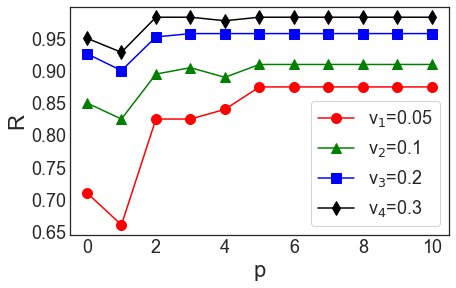}
    \vspace{-0.5cm}
    \label{fig:14AC-DR-p4}
  \end{subfigure}
  \begin{subfigure}{0.241\textwidth}
    \includegraphics[width=\linewidth]{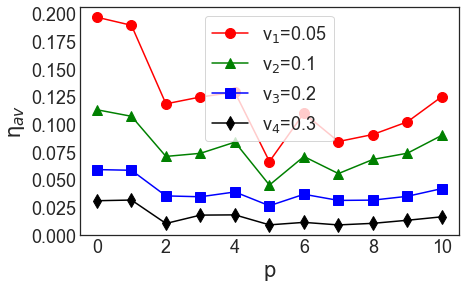}
    \vspace{-0.6cm}
    \label{fig:14AC-AT-p4}
  \end{subfigure}
  \caption{The performance of MTD-strengthened DNN over the number of adversarially trained models (14-bus AC, $K=10$)}
  \label{fig:14ACp}
  \vspace{-0.2cm}
\end{figure}

We also perform simulations under the AC power flow model, following a methodology similar to that employed in the DC condition. Figure~\ref{fig:14ACp} denotes the performance of MTD-strengthened DNN according to $p$. Similar to the trend under DC conditions, defense is more efficient with larger $p$ values.

\begin{figure}[t]
  \centering
  \begin{subfigure}{0.241\textwidth}
    \includegraphics[width=\linewidth]{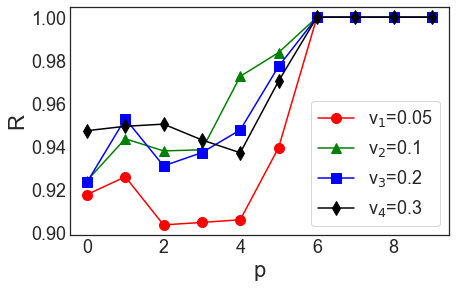}
    \vspace{-0.6cm}
    \caption{IEEE 30-bus DC}
    \label{fig:30DC-DR-p4}
  \end{subfigure}
  \begin{subfigure}{0.241\textwidth}
    \includegraphics[width=\linewidth]{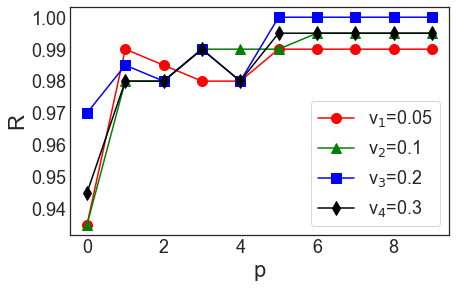}
    \vspace{-0.6cm}
    \caption{IEEE 118-bus DC}
    \label{fig:118DC-DR-p4}
  \end{subfigure}
  \caption{The recall rate of MTD-strengthened DNN over the number of adversarially trained models ($K=10$)}
  \label{fig:DCA30118}
  \vspace{-0.5cm}
\end{figure}

{\bf Simulations With Large Bus Systems:}
We conducted simulations on IEEE 30 and 118-bus systems to demonstrate the scalability of our solution to large bus systems. In Fig~\ref{fig:DCA30118}, we plot the recall rate of the MTD-strengthened DNN against adversarial FDIAs as a function of $p$. We observe the similar detection performance to that of the IEEE 14-bus system.

\begin{figure}[t]
	\centering
	\includegraphics[width=0.45\textwidth]{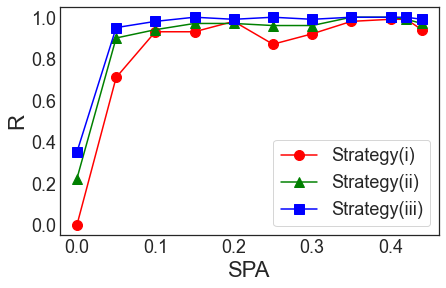}
    \vspace{-0.25cm}
	\caption{Performance of physics-based MTD integration strategies over SPA.}
    \vspace{-0.2cm}
	\label{fig:14DC-SPA}
\end{figure}

%\arrayrulecolor{blue}

\begin{comment}

\begin{table}[t]
	\centering
 \caption{Comparison of MTD strategies}
 \resizebox{0.49\textwidth}{!}{
    	\begin{tabular}{|l|c|c|c|}
        \hline
         \makecell[c]{Strategy} & Recall rate & Time (s) & \makecell{Increase in \\ OPF costs}\\
		\hline
        MTD-strengthened DNN ($p=6, K=10$) &  $0.942$ & $3832$ & 0\\
		\hline
        BDD strengthened using Physics-based MTD ($\text{SPA}=0.4$) & $>0.99$ & 0 & 1.6\%\\
		\hline
        \makecell[l]{S(i): DNN strengthened using Physics-based MTD ($\text{SPA}=0.4$)} &  $>0.99$ & $19$ & 1.6\%\\
		\hline
       \makecell[l]{S(ii): MTD-strengthened DNN ($p=0, K=10$) \\ \hfill + Physics-based MTD ($\text{SPA}=0.35$)} & $>0.99$ & $1092$ & 0.96\% \\
		\hline
       \makecell[l]{S(iii): MTD-strengthened DNN ($p=1, K=10$) \\ \hfill + Physics-based MTD ($\text{SPA}=0.15$)} & $>0.99$ & $1551$ & 0.1\% \\
		\hline
	\end{tabular}
 }
	\label{tbl: Comparison of MTD approaches}
 \vspace{-0.3cm}
\end{table}

\end{comment}

\begin{table}[t]
	\centering
 \caption{Comparison of MTD strategies}
 \resizebox{0.49\textwidth}{!}{
    	\begin{tabular}{|l|c|c|c|}
        \hline
         \makecell[c]{Strategy} & Recall rate & Time (s) & \makecell{Increase in \\ OPF costs}\\
		\hline
        MTD-strengthened DNN ($p=6, K=10$) &  $0.942$ & $3832$ & 0\\
		\hline
        BDD strengthened using Physics-based MTD ($\text{SPA}=0.4$)& $>0.99$ & 0 & 1.6\%\\
		\hline
        \makecell[l]{S(i): DNN strengthened using Physics-based MTD ($\text{SPA}=0.4$)} &  $>0.99$ & $19$ & 1.6\%\\
		\hline
       \makecell[l]{S(ii): MTD-strengthened DNN ($p=0, K=10$) \\ \hfill + Physics-based MTD ($\text{SPA}=0.35$)} & $>0.99$ & $1092$ & 0.96\% \\
		\hline
       \makecell[l]{S(iii): MTD-strengthened DNN ($p=1, K=10$) \\ \hfill + Physics-based MTD ($\text{SPA}=0.15$)} & $>0.99$ & $1551$ & 0.1\% \\
		\hline
	\end{tabular}
 }
	\label{tbl: Comparison of MTD approaches}
 \vspace{-0.2cm}
\end{table}

\begin{figure}[t]
	\centering
	\includegraphics[width=0.5\textwidth]{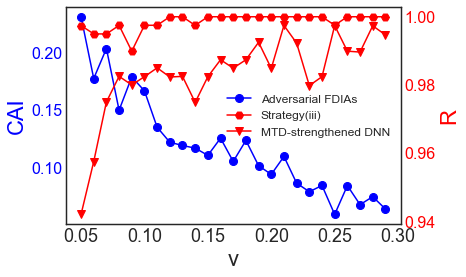}
    \vspace{-0.2cm}
	\caption{Performance of strategy(iii) on different adversarial FDIA test cases.}
	\label{fig:CAI-R}
 \vspace{-0.2cm}
\end{figure}

{\bf Integration of Physics-Based MTD:} Finally, we investigate the integration of MTD-strengthened DNN with physics-based MTD. We follow the approach detailed in \cite{LakshminarayanaCostBenefit} to implement the physics-based MTD. For IEEE 14-bus system, the D-FACTS devices are installed on 7 branches indexed by $\mathcal{L}_D = \{1,5,9,11,14,17,19\}$, and the branch flow limits are set to be 160 MWs for link 1 and 60 MWs for all other links of the power system.
%, in which \chen{the SPA is applied to assess MTD effectiveness, and the increase in OPF cost is applied to quantify operational costs.} {\color{red}[REMOVE] the MTD reactance perturbations are designed to increase the SPA between the column spaces of $\Hm$ and $\Hm^{\prime}$. Specifically, MTD reactance perturbations resulting in higher SPA are more effective in detecting stealthy FDI attacks. MTD designed according to this metric is shown to have robust performance in the presence of sensor measurement noise \cite{xu2021robust} as well as achieve effective FDI detection under AC conditions \cite{LakshminarayanaCostBenefit}.} 
The optimal reactance perturbation is solved in MATLAB using Sequential Quadratic Programming (SQP) via \emph{fmincon}, which is a gradient-based deterministic solver for constrained nonlinear optimization. Additionally, the MultiStart metaheuristic is applied to enhance the global search by running the optimization from multiple starting points. Then, we integrate physics-based MTD following the method described in Section~\ref{Physics-based MTD} (also illustrated in Figure~\ref{fig:DNN physic MTD}). Specifically, we consider three strategies -- (i) implementing physics-based MTD and adaptation of the base model from $f_b (\omegav_b)$ to $f^{\prime}_b (\omegav^{\prime}_b)$ only, (ii) creating student models from $f^{\prime}_b (\omegav^{\prime}_b)$ with $p=0, K=10$, and (iii) creating student models from $f^{\prime}_b (\omegav^{\prime}_b)$ with $p=1, K=10$. All strategies are tested using the testing dataset $\nu_1$, which is defined in Section~\ref{sec: Simulation Results}. 

The simulation results are presented in Figure~\ref{fig:14DC-SPA}, which shows the recall rate of the three strategies as a function of the SPA used to implement the physics-based MTD. Additionally, Table~\ref{tbl: Comparison of MTD approaches} compares recall rates and the execution time to implement the schemes. 
It can be observed that the recall rate of all strategies increases by increasing the SPA, exceeding $99\%$ when SPA is larger than $0.4$. Moreover, Strategy~(i) achieves an improved recall rate compared to applying MTD-strengthened DNN alone (Table~\ref{tbl: Comparison of MTD approaches}), thus showing the effectiveness of combining physics and MTD-strengthened DNN approach. However, note that combining with the physics-based approach incurs operational costs, which, in turn, increases as we implement physics-based MTD with higher SPA.  It can be observed that Strategy~(iii) achieves a recall rate of over $99\%$ 
with a SPA of $0.15,$ with $p = 1$ (i.e., with just one adversarially trained model). Thus, it achieves a balance between keeping the operational costs and computational costs at low values. 

% with a balanced time consumption of $1551$ seconds and relatively low operational costs of $\text{SPA} = 0.15$. This makes it superior to other strategies that either consume significantly more time, incur high operational costs, or do not achieve a high recall rate. These results indicate that properly integrating physics-based MTD with MTD-strengthened DNN can achieve a cost-benefit balance.

% However, implementing physics-based MTD incurs unavoidable operational costs due to deviations in optimal generation dispatch, and a larger SPA corresponds to higher operational costs. In contrast, the operational costs of MTD-strengthened DNN are related to its requirements for computational resources, memory, and time consumption, which can be reduced with technological advancements. Therefore, combining physics-based MTD and MTD-strengthened DNN is reasonable for balancing defense effectiveness and operational costs.

% Additionally, it can be observed in Figure~\ref{fig:14DC-SPA} that strategy(ii) can achieve an improved recall rate (up to $8\%$) compared to strategy(i). However, a large SPA (i.e., $\text{SPA} \geq 0.35$) is still needed for the recall rate of this strategy to exceed $99\%$. Therefore, strategy(ii) does not significantly reduce the overall operational costs.

% Alternatively, the strategy(iii) further improves the defense performance. We summarize the settings for each strategy to achieve its maximum recall rate in Table~\ref{tbl: Comparison of MTD approaches}.

{\bf Simulations With Different Adversarial FDIAs:} Our proposed strategies have proven effective on the testing dataset $\nu_1$, which is composed of adversarial FDIAs aimed at hiding BDD-bypassing FDIAs with a magnitude of $\nu=0.05$. We further test our strategy on adversarial FDIAs designed to hide other magnitudes of BDD-bypassing FDIA. Specifically, we test using Strategy~(iii) and the MTD-strengthened DNN with the same settings as in Table~\ref{tbl: Comparison of MTD approaches}. We applied the metric of Change of Attack Intensity (CAI) to assess the influence of adversarial perturbations on the original attack target of BDD-bypassing FDIAs. CAI, originally introduced in \cite{AFDIA2021}, is defined as the ratio of the attack magnitude ($L_2$ norm) before and after an adversarial attack, as given by $
    CAI = \frac{||\av + \deltav||_2}{||\av||_2}.$    
A CAI value close to 1 indicates minimal influence of the adversarial perturbation on the attack vector of the BDD-bypassing FDIA it aims to hide.

The simulation results, presented in Figure~\ref{fig:CAI-R}, illustrate the CAI of the adversarial FDIAs and the recall rate of Strategy~(iii) and MTD-strengthened DNN as functions of $\nu$. It can be observed that the CAI of the adversarial FDIAs decreases as $\nu$ increases. This indicates that when attackers aim to hide larger BDD-bypassing FDIAs, they need to either alter the original attack vector more significantly or reduce the attack magnitude. Additionally, the recall rate of Strategy~(iii) and the MTD-strengthened DNN increases with $\nu$. This suggests that when the magnitudes of BDD-bypassing FDIAs are larger, the developed adversarial FDIAs more likely to be detected. Furthermore, the recall rate of Strategy~(iii) remains over 99\% for all test cases. This result further validates the effectiveness of Strategy~(iii) under different adversarial FDIA settings.

{\bf Comparison with the State-of-the-Art:}
We also provide a comparison of our approach with model-based methods (i.e., physics-based MTD) and ML-based defense methods. The results are presented in Table~\ref{tab: Comparison of mainstream techniques}. For ML-based defense against adversarial attacks, we have compared our approach with several mainstream methods, including four static defenses (i.e., defensive distillation, gradient masking, adversarial training on FGSM attacks, and adversarial training on CW attacks) and two dynamic defenses (i.e., randomization of model parameters and ensemble methods such as fMTD \cite{fMTD2019}). The results show that static defenses fail to defend against adaptive attackers who continuously probe the latest defense settings (e.g., model parameters); as such, attackers can consistently generate feasible A-FDIAs that bypass the defense (resulting in a detection accuracy of 0). This observation is consistent with the results of prior work on MTD applied in the context of image processing task \cite{Morphence2021}. On the other hand, dynamic defenses, such as randomization of model parameters or ensemble methods, cannot achieve reliable defense against A-FDIAs while also suffering from low accuracy in identifying legitimate measurements and traditional FDIAs. Additionally, applying only physics-based MTD is ineffective in defense when the SPA value is low (e.g., SPA = $0.15$). A sufficiently large SPA (e.g., SPA = $0.4$) is required for effective defense, but this leads to high operational costs. For more details, please refer to Table V. In contrast, our proposed method achieves high detection accuracy across all test cases even with a low SPA value.

%\renewcommand{\arraystretch}{0.9} % Adjust row height
%\arrayrulecolor{blue}
\begin{table}[t]
    \centering
    \caption{Comparison of mainstream techniques}
    \resizebox{0.5\textwidth}{!}{
    \begin{tabular}{cccccccc}
    \toprule
    \hline
        \multicolumn{2}{c}{Method} & \multicolumn{2}{c}{IEEE 14-bus} & \multicolumn{2}{c}{IEEE 30-bus} & \multicolumn{2}{c}{IEEE 118-bus} \\
        \cmidrule(r){3-4} 
        \cmidrule(r){5-6}
        \cmidrule(r){7-8}
        \vspace{-5mm}
        & & \renewcommand{\arraystretch}{0.5} A-FDIAs \vspace{0.1cm} & \renewcommand{\arraystretch}{0.5} \begin{tabular}[c]{@{}c@{}}Legitimate\\ measurements\\ \& FDIAs \end{tabular} \vspace{0.1cm} & \renewcommand{\arraystretch}{0.5} A-FDIAs \vspace{0.1cm} & \renewcommand{\arraystretch}{0.5} \begin{tabular}[c]{@{}c@{}}Legitimate\\ measurements\\ \& FDIAs \end{tabular} \vspace{0.1cm} & \renewcommand{\arraystretch}{0.5} A-FDIAs \vspace{0.1cm} & \renewcommand{\arraystretch}{0.5} \begin{tabular}[c]{@{}c@{}}Legitimate\\ measurements\\ \& FDIAs \end{tabular} \vspace{0.1cm} \\
        \hline
        \multirow{4}{*}{\renewcommand{\arraystretch}{0.5} \begin{tabular}[c]{@{}c@{}}Static\\ Defense\end{tabular}} & Defensive Distillation & $0$ & $0.969$ & $0$ & $0.974$ & $0$ & $0.945$ \\
        \cline{2-8}
        & Gradient masking & $0$ & $0.972$ & $0$ & $0.962$ & $0$ & $0.979$\\
        \cline{2-8}
        & \renewcommand{\arraystretch}{0.5} \begin{tabular}[c]{@{}c@{}}Adversarial training\\ on FGSM attacks\end{tabular} & $0$ & $0.895$ & $0$ & $0.924$ & $0$ & $0.948$\\
        \cline{2-8}
        & \renewcommand{\arraystretch}{0.5} \begin{tabular}[c]{@{}c@{}}Adversarial training\\ on CW attacks\end{tabular} & $0$ & $0.989$ & $0$ & $0.984$ & $0$ & $0.913$\\
        \hline
        \multirow{2}{*}{\renewcommand{\arraystretch}{0.5} \begin{tabular}[c]{@{}c@{}}Dynamic\\ Defense\end{tabular}} & \renewcommand{\arraystretch}{0.5} \begin{tabular}[c]{@{}c@{}}Randomization of\\ model parameters\end{tabular} & $0.880$ & $0.968$ & $0.918$ & $0.978$ & $0.798$ & $0.969$ \\
        \cline{2-8}
        & \renewcommand{\arraystretch}{0.5} \begin{tabular}[c]{@{}c@{}}Ensemble method\\ (e.g. fMTD \cite{fMTD2019})\end{tabular} & $0.844$ & $>0.99$ & $0.723$ & $>0.99$ & $0.915$ & $>0.99$\\
        \hline
        \renewcommand{\arraystretch}{0.5} \begin{tabular}[c]{@{}c@{}}Physics-based\\MTD \end{tabular} & SPA = 0.15 & $0.09$ & $0.582$ & $0.02$ & $ 0.214 $ & $ 0.05$ & $0.339$ \\
        \hline
        \renewcommand{\arraystretch}{0.5} \begin{tabular}[c]{@{}c@{}}Physics-based\\MTD +ML \end{tabular} & Our approach & $>0.99$ & $>0.99$ & $>0.99$ & $>0.99$ & $>0.99$ & $>0.99$ \\
        \hline
    \bottomrule    
    \end{tabular}
    }
    \label{tab: Comparison of mainstream techniques}
    \vspace{-0.5cm}
\end{table}
\arrayrulecolor{black}

\vspace{-3mm}
\subsection{Key Findings}
(i) The results in Figures~\ref{fig:14DCn}, \ref{fig:14DCp}, \ref{fig:14ACp}, and \ref{fig:DCA30118} demonstrate that MTD-strengthened DNNs can achieve moderate level of accuracy. The detection accuracy improves with an increase in the number of models in the pool and the number of adversarially trained models, but plateaus beyond a certain threshold. Notably, the average transferability rate of adversarial attacks decreases. This confirms that the effectiveness of MTD-strengthened DNNs lies in their ability to reduce the transferability of adversarial attacks among models in the pool.
The results in Figure~\ref{fig:DCA30118}, Table~\ref{tbl:execution times}, and Table~\ref{tbl:time consumption} demonstrate the effectiveness of MTD-strengthened DNNs in large bus systems. (ii) Furthermore, applying MTD-strengthened DNNs does not significantly increase execution time compared to using a single DNN model. However, while increasing the number of adversarially trained models improves detection performance, it also raises computational costs during offline training. This creates a trade-off that must be considered when configuring the hyperparameters of MTD-strengthened DNNs.
(iii) The results in Figure~\ref{fig:14DC-SPA}, \ref{fig:CAI-R}, and Table~\ref{tbl: Comparison of MTD approaches} demonstrate that integrating physics-based MTD with MTD-strengthened DNNs can significantly improve detection performance, achieving accuracy exceeding 99\%. This integration also reduces the number of adversarially trained models required, thereby lowering computational costs. Furthermore, this integration results in minimal increases in OPF cost compared to using physics-based MTD to strengthen either BDD or a single DNN.

\vspace{-3mm}
\section{Conclusions} \label{Conclusion}
This study has investigated defending against the threat of adversarial FDIAs in power grid state estimation. We propose an MTD-strengthened DNN approach, which creates a MTD model pool instead of deploying a static DNN model, such the transferability of adversarial FDIAs within the model pool is low. Furthermore, we propose to improve the MTD performance by combining it with a physics-based MTD approach. The simulation results show that combining the two techniques can achieve very high detection accuracy while keeping the MTD's operational and computational costs low. The proposed defence is sensitive to the selection of hyperparameters, which should be carefully chosen according to practical power grid conditions. This study shows that incorporating the concept of MTD can effectively defend against adversarial FDIAs in power grids. 

%\chen{Future work will focus on developing analytical metrics to design effective MTD strategies against adversarial FDIAs in power grids.}

%\chen{Future research could focus on evaluating performance on real-world, large-scale smart grids, and addressing practical challenges such as communication delays, real-time data acquisition, and system coordination, developing robust and adaptive MTD mechanisms that evolve in response to emerging cyberattacks.}

%\chen{Future work could focus on enhancing the MTD approach by enabling attack localization to identify compromised sensors or communication links. Additionally, deploying the proposed framework on real-world, large-scale smart grids introduces challenges related to communication delays, data acquisition intervals, and handling noisy or incomplete data. Addressing these issues may involve integrating event-triggered mechanisms, improving fault tolerance, and optimizing communication and data acquisition protocols. Moreover, incorporating theoretical performance guarantees through physics-informed machine learning and developing adaptive game-theoretic defenses could further enhance resilience against evolving cyberattacks.}

Building on this work, there are several interesting future research directions. 
First, while the proposed MTD approach
is aimed at detecting adversarial FDI attacks with high accuracy, it does not localize the attacks, i.e., pinpoint the sensors/communication links that are the target of the attacker, which can be an important area of improvement. 
This also relates to interpretability or explainability issues in ML models. Second, testing the approach on real-world, large-scale grids while addressing challenges like data outliers, communication delays, real-time data acquisition, and system coordination would be valuable. To this end, using robust feature extraction approaches that can reconstruct the information in noisy/outlier datasets will be useful. Finally, developing robust and adaptive MTD mechanisms that can evolve in response to emerging cyberattacks would further improve the system's resilience. To this end, adopting game-theoretic approaches can be a promising future research direction.

\vspace{-3mm}
\bibliographystyle{IEEEtran}
\bibliography{IEEEabrv,bibliography}

\end{document}